\documentclass[10pt,conference]{IEEEtran}
\usepackage{cite}
\usepackage{amsmath,amssymb,amsfonts}
\usepackage{algorithmic}
\usepackage{graphicx}
\usepackage{textcomp}
\usepackage{xcolor}
\usepackage{multirow}
\usepackage[hyphens]{url}
\usepackage{fancyhdr}
\usepackage{hyperref}
\usepackage{cleveref}
\usepackage{tcolorbox}
\usepackage[table]{xcolor}
\usepackage{enumitem}
\usepackage{booktabs}
\usepackage{tikz}
\usepackage{balance}          % n

\newcommand{\ignore}[1]{}

\newcommand{\fix}[1]{{\color{blue}#1}}

\definecolor{advpink}{RGB}{250,218,221}
\definecolor{advpinkline}{RGB}{201,86,96}
\definecolor{noadvgreen}{RGB}{219,239,219}
\definecolor{noadvgreenline}{RGB}{90,150,90}

\newcommand{\hpcayear}{2027}

\title{From Fleet to Lab: Revisiting the  Security and Complexity of Industrial Rowhammer Mitigation}

\newcommand\hpcaauthors{First Author$\dagger$ and Second Author$\ddagger$}
\newcommand\hpcaaffiliation{First Affiliation$\dagger$, Second Affiliation$\ddagger$}
\newcommand\hpcaemail{Email(s)}

\author{
  \ifdefined\hpcacameraready
    \IEEEauthorblockN{\hpcaauthors{}}
      \IEEEauthorblockA{
        \hpcaaffiliation{} \\
        \hpcaemail{}
      }
  \else
    \IEEEauthorblockN{Hritvik Taneja}
    \IEEEauthorblockA{
      Georgia Institute of Technology \\
      htaneja3@gatech.edu
    }
    \and
    \IEEEauthorblockN{Moinuddin Qureshi}
    \IEEEauthorblockA{
      Georgia Institute of Technology \\
      moin@gatech.edu
    }
  \fi 
}

\fancypagestyle{camerareadyfirstpage}{%
  \fancyhead{}
  
  \fancyhead[C]{
    \ifdefined\aeopen
    \parbox[][12mm][t]{13.5cm}{\hpcayear{} IEEE International Symposium on High-Performance Computer Architecture (HPCA)}    
    \else
      \ifdefined\aereviewed
      \parbox[][12mm][t]{13.5cm}{\hpcayear{} IEEE International Symposium on High-Performance Computer Architecture (HPCA)}
      \else
      \ifdefined\aereproduced
      \parbox[][12mm][t]{13.5cm}{\hpcayear{} IEEE International Symposium on High-Performance Computer Architecture (HPCA)}
      \else
      \parbox[][0mm][t]{13.5cm}{\hpcayear{} IEEE International Symposium on High-Performance Computer Architecture (HPCA)}
    \fi 
    \fi 
    \fi 
    \ifdefined\aeopen 
      \includegraphics[width=12mm,height=12mm]{ae-badges/open-research-objects.pdf}
    \fi 
    \ifdefined\aereviewed
      \includegraphics[width=12mm,height=12mm]{ae-badges/research-objects-reviewed.pdf}
    \fi 
    \ifdefined\aereproduced
      \includegraphics[width=12mm,height=12mm]{ae-badges/results-reproduced.pdf}
    \fi
  }
  \fancyfoot[C]{}
}
\begin{document}
\maketitle

%Enables the camera ready header and footer
\ifdefined\hpcacameraready 
  \thispagestyle{camerareadyfirstpage}
  \pagestyle{empty}
\else
  \thispagestyle{plain}
  \pagestyle{plain}
\fi

\newcommand{\hpcaheight}{0mm}
\ifdefined\eaopen
\renewcommand{\hpcaheight}{12mm}
\fi

%%%%%%%%%%%%%%%%%%%%%%%%%%%%%%%%%%%%%%%%
%%%%%%%% -- PAPER CONTENT STARTS -- %%%%%%%%%

\ignore{
FLOW:

- Rowhammer MC side 
- Storage and Mitig tradeoff
- Filtered Rowhammer mitigation
- Sigries (what it is), six crieterion
- We analyze: weak security, CAM, storage

In this paper, we do analysis and show that we can get security, avoid CAM and still meet the six crieterion.  Insight, do not optimize for both independently, cooptimize.  Use simple SRAM filters instead of CAM, which gives zero overheads for benign workloads.  The overhead for stressful application can be the same as Segries by ensuring security for the epoch in which switching occurs, and then having multiple epocs with heavy mode.  

We also show that the principles apply to deterministic mitigations as well. We delveop FIRM-D, which has half as much hardware, no CAM, no reliance on probablistic mitigations, ensures 0 slowdown for benign and the given slowdown under stressfull pattern.s 

}

%Sigries is shown to fulfill all six criteria outlined in the paper (minimal slowdown when not under attack, no performance outliers, liveness, storage, flexibility, and configurable security guarantees). 

%In this paper, we show that all six criteria outlined in the Sigries paper can be met while providing guaranteed security,  without requiring any CAM, and at less than half the storage overheads. 

\begin{abstract}

This paper studies efficient and secure Rowhammer mitigation at the
Memory-Controller (MC). Rowhammer mitigation faces a fundamental tradeoff
between tracking storage and mitigation rate: precise trackers (such as
Misra-Gries) avoid unnecessary mitigations but require large CAM
structures, whereas sampling-based schemes (such as PARA) require no
storage but incur frequent mitigations even when not under attack.
Microsoft recently deployed Sigries, an MC-side Rowhammer defense that
combines an under-provisioned Misra-Gries tracker with a row-sampling
fallback, in its Azure Cobalt 200 SoC. Sigries observed that the
tracker-to-sampling transition can be insecure, and claimed the reverse
transition is always safe. Our analysis shows that this transition is
also vulnerable, and that a \emph{Round-Robin Attack} across sub-banks
reduces the Mean-Time-To-Failure of Sigries to about one second, eight
orders of magnitude below the 13 years with PARA. Sigries also suffers from
CAM complexity and high storage overheads. Our goal is to develop a
solution that is fully secure and minimizes the storage and complexity of
Sigries, while matching its performance.

Our proposal, {\em FiRM (Filtered Rowhammer Mitigation)}, is based on the key insight that for a secure design the tracking-mode and sampling-mode should not be configured independently, but co-designed to ensure that the system remains secure not only during the two modes but also during the two transitions. FiRM also avoids the complexity of Sigries by replacing the CAM-based tracker with simple SRAM filters. As benign workloads do not exceed the filtering threshold, FiRM incurs no slowdown for benign workloads. To handle stressful patterns, we propose two variants.  First, {\em FiRM-P (probabilistic)}, which relies on PARA; however, it uses varying probabilities during transitions and steady state to ensure both security and low performance overhead.  FiRM-P is secure, has less than half the storage overhead of Sigries, and provides similar performance.  Second, {\em FiRM-D (deterministic)}, which provides guaranteed deterministic security by modulating the rate of mitigation (zero for benign workloads and non-zero under attack). FiRM-D requires less than two-thirds the storage of Sigries, does not require any CAM, and has zero slowdown for benign workloads.  Our paper shows that a principled approach can avoid both the insecurity and the complexity of Sigries.

\end{abstract}

\begin{figure*}[!htb]
    \centering
\includegraphics[width=6.9in]{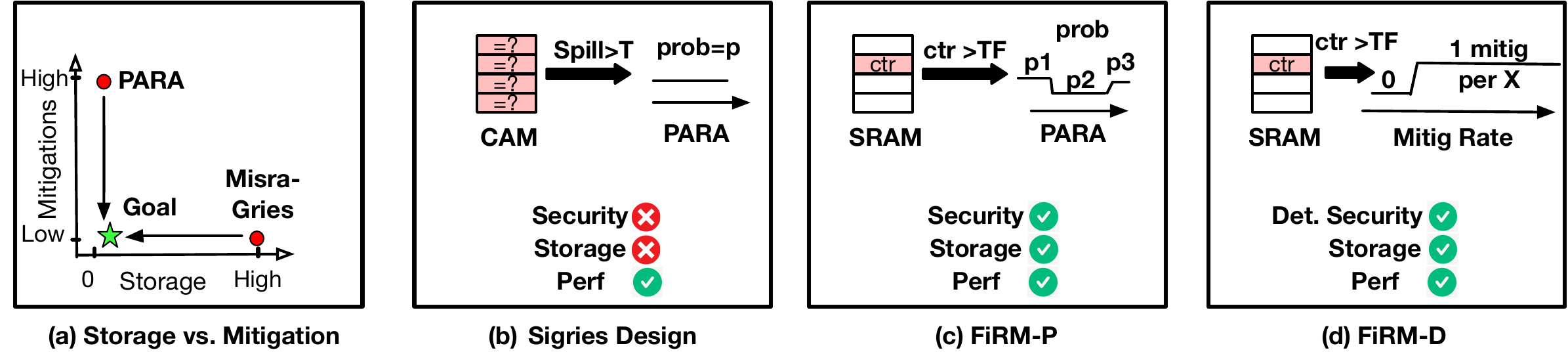}
\vspace{-0.1 in}
    \caption{  (a) Counter-based trackers need more storage, whereas PARA needs more frequent mitigations; ideally, we want low storage and low mitigation.  (b) Sigries uses a per-sub-bank shortened Misra-Gries tracker (CAM)  and a uniform rate for PARA as fallback; however, it suffers from both insecurity and complexity. (c) Our proposal, FiRM-P, provides principled Filtered Rowhammer Mitigation and ensures security in both modes and transitions, and avoids CAM complexity by using simple SRAM filters. (d) Our proposal, FiRM-D, provides deterministic security by modulating the mitigation rate of a region.  }
\vspace{-0.2 in}
    \label{fig:intro}
\end{figure*}

\section{Introduction}

%Rowhammer occurs when a row is frequently activated, leading to charge loss in neighboring rows.  The number of activations required to cause a bitflip is called the {\em Rowhammer Threshold (TRH).}  Over the last decade, as the sizes of the DRAM cells have reduced, the TRH has reduced from 135K~\cite{kim2014flipping} to 4.8K~\cite{kim2020revisitingRH}. Rowhammer provides an attacker with a powerful tool to flip bits in critical data structures, such as page tables, and cause privilege escalation. Thus, Rowhammer is not only a reliability concern, but a serious security threat~\cite{seaborn2015exploiting,gruss2018another,frigo2020trrespass,zhang2020pthammer}\cite{kwong2020rambleed}. Hardware mitigation of Rowhammer typically relies on a tracking mechanism to identify the {\em aggressor} rows and then refreshing a specific number of {\em victim} rows on either side of the aggressor row. As in-DRAM mitigations can transparently handle Rowhammer in the DRAM chips, we focus on in-DRAM mitigations. 

DRAM scaling has enabled denser chips  but it has also
made DRAM increasingly susceptible to data-disturbance errors, such as
Rowhammer~\cite{kim2014flipping}. Rowhammer occurs when frequent
activations to a DRAM row cause bit-flips in nearby rows. Rowhammer is not
merely a reliability concern but a serious security threat~\cite{seaborn2015exploiting,kwong2020rambleed}.
The severity of Rowhammer is characterized by the {\em Rowhammer Threshold}
($T_{RH}$), the minimum number of activations required to induce a bit-flip.
%The $T_{RH}$ has dropped from
%140K~\cite{kim2014flipping}~(2014) to 4.8K~\cite{kim2020revisitingRH}~(2020).
 
\noindent{\bf Why Memory-Controller Side Mitigation?}
A typical hardware Rowhammer defense has two parts: a {\em tracker} that
identifies aggressor rows, and a {\em mitigation} that refreshes the victim rows. Mitigation can be done either inside the DRAM (in-DRAM)
or at the Memory Controller (MC). Commercially deployed in-DRAM trackers such as TRR have been broken~\cite{frigo2020trrespass,jattke2021blacksmith,hassan2021UTRR}. Principled in-DRAM solutions such as {\em Per-Row Activation Counting (PRAC)} incur significant slowdown (about 8.4\%) even when the system is not under an attack.  MC-based mitigations can provide strong security while incurring lower overheads.  Furthermore, recent trends in hyperscalers~\cite{gholkar2026vistara} to reuse DIMMs from decommissioned servers mean that currently deployed DDR5 DIMMs (insecure against Rowhammer) may remain in use for more than a decade, further boosting the need for MC-side Rowhammer defense.

%JEDEC DDR5 exposes {\em Directed Refresh Management (DRFM)}, which lets the MC specify an aggressor row and trigger victim-refresh without knowing the internal address mapping of the DRAM. 

\vspace{0.05 in}
\noindent{\bf The Storage vs. Mitigation Tradeoff:}
MC-side trackers fall into two classes. {\em Deterministic} (counter-based) trackers, such as
Graphene~\cite{park2020graphene}, use the Misra-Gries algorithm to identify the
top-$K$ most activated rows. As these trackers  issue mitigations only when needed, benign workloads experience virtually zero mitigations. However, they require a CAM structure with hundreds of entries (per bank), making them impractical for commercial adoption. At the other end, {\em
probabilistic} (sampling-based) trackers, such as PARA~\cite{kim2014flipping}, select a row for mitigation with probability $p$. While they do not need any storage, they issue frequent mitigations 
even when the system is not under an attack. Ideally, we want the mitigation-efficiency of counter-based trackers and storage-efficiency of sampling-based trackers, as shown in Figure~\ref{fig:intro}(a).

\vspace{0.05 in} 
\noindent{\bf Sigries:} Recently, Microsoft deployed {\em Sigries}~\cite{sigries} (in Azure Cobalt 200),
an MC-side Rowhammer defense that targets the above trade-off. Sigries splits the bank into several sub-banks, and equips each sub-bank with an {\em
under-provisioned} Misra-Gries tracker (containing a couple of dozen entries) and row-sampling fallback, as shown in Figure~\ref{fig:intro}(b). The tracker is sufficient to handle
benign workloads without triggering mitigations, thus benign workloads incur zero slowdown.  When an access pattern
 overflows the tracker, the design falls back to PARA, providing probabilistic protection at negligible additional storage. As
the tracker is under-provisioned, each sub-bank of Sigries needs far fewer CAM entries
than per-bank Graphene. However, Sigries still incurs the complexity of a CAM lookup of the sub-bank entries on each activation. 

% \newpage
The two components of Sigries (Misra-Gries and PARA) are each designed independently to be secure within their mode.  However, combinations of two things that are independently secure can still be insecure. Sigries identifies that the transition from tracker to sampling can be unsafe and handles this by limiting the vulnerability windows to less than 1 hour per year (per sub-bank). It does so by running the sampling-mode for several minutes on each transition.  Our paper investigates whether Sigries has any additional unknown vulnerabilities.

%Sigries claims that the transition from sampling-mode to tracker-mode is {\em always} safe, so no additional provisioning is needed to handle this transition. 

%This is important as Sigries is not a theoretical design but something deployed in commercial systems. 

\ignore{
\vspace{0.05 in}
\noindent{\bf Questions:} This paper investigates the following questions:
\vspace{0.02 in}

{\bf (1)} Does the Sigries design have any additional undisclosed vulnerabilities? This is important as Sigries is not just a theoretical design but something deployed in commercial systems. 

{\bf (2)} What is the impact of Sigries vulnerability target (of one hour per year) on the mean-time-to-failure of the system? 

{\bf (3)} Can we design an MC-side mitigation that provides strong security (no vulnerability windows), avoids CAM complexity, and yet meets the performance of Sigries?

{\bf (4)} Is it possible to design a dual-mode scheme that is fully deterministic and thus avoids any probabilistic fall-back (and associated vulnerabilities from random-number generation~\cite{soothsayer})?  
}

\vspace{0.05 in}

While the Sigries paper asserts that the transition from sampling-mode to tracker-mode is {\em always} safe, our analysis shows otherwise. Our key observation is that tracker-mode is safe only if the row started with zero unmitigated activations, however, for Sigries, the row would have hundreds of unmitigated activations (due to prior sampling-mode), making the transition unsafe.  We also show that while the target vulnerability of one hour per year (per sub-bank) may seem low, an attack that {\em Round-Robins} multiple sub-banks can cause {\bf a failure within a second}, reducing the system MTTF by eight orders of magnitude. The \textbf{goal} of this paper is to design a secure mitigation that also avoids the complexity of Sigries.

\vspace{0.05 in}
 
\noindent{\bf Insight and Solution.}
The root cause of the vulnerability of Sigries is that the two solutions (tracker and sampling) are independently designed to be secure in isolation, assuming the given row starts with zero activations. When the two schemes are combined, this assumption gets violated, which causes unsafe behavior.  The key insight of our paper is that to ensure security both modes should be co-designed to ensure that not only are they secure in isolation but also during the transition from one to another. Our proposal,
{\bf{FiRM (Filtered Rowhammer Mitigation)}}, is built on this insight.  FiRM co-designs the filter and the fallback so that the activation budget of an attacker is bounded across all modes and transitions.

We also observe that the filter does not 
need the precision of Misra-Gries, it simply needs to track if the benign traffic to the given region is below a certain threshold.
FiRM therefore replaces the CAM-based tracker of Sigries with a simple {\em
SRAM filter}: an untagged, direct-mapped table of counters that
requires no associative lookup. As benign workloads do not exceed the
{\em Filtering Threshold} ($T_F)$, FiRM incurs {\em zero} slowdown on benign workloads, while
relying on the fallback under attacks. We propose two
variants that differ in fallback.

Our first design, {\bf FiRM-P (Probabilistic)},
uses PARA as its fallback, however, unlike Sigries, it does not use a
uniform probability for mitigation. The activations an attacker can accrue while crossing a mode
boundary are larger than those accrued in steady state, so FiRM-P uses an
elevated probability during the transition windows and relaxes to a lower
steady-state probability during steady-state, as shown in Figure~\ref{fig:intro}(c). This lets FiRM-P avoid security vulnerabilities during transitions while still ensuring similar steady-state performance to Sigries during sampling mode. FiRM-P is secure by construction, avoids CAM complexity, requires less than half the storage, and has similar performance as Sigries.
 
Our second design, \noindent{\bf FiRM-D (Deterministic),} avoids reliance on random number generator (RNG),  which can be  
a source of both design complexity and insecurity~\cite{soothsayer}.  Rather
than sampling, FiRM-D modulates the {\em rate} of mitigation of a region based on the activation count. As shown in Figure~\ref{fig:intro}(d), the rate is zero when the filter is below $T_F$ (benign
workloads) and non-zero thereafter to ensure security for the attacked row. FiRM-D provides
strong deterministic security, uses no CAM or RNG, requires less than two-thirds the storage of Sigries, and has zero slowdown for benign workloads.

\vspace{0.05 in}

\noindent{\bf Contributions:} Our paper makes the following contributions:
\begin{enumerate}
    \item We show that Sigries has a previously unidentified vulnerability during the
    heavy-to-lite transition. We also show that under multi-sub-bank attack the MTTF of Sigries reduces to 1 second, instead of 13 years for PARA. 
 
    \item We propose {\em FiRM}, a principled filtered mitigation, that configures
     the tracking and sampling components 
    jointly, so that security holds across both modes
    and transitions.
 
    \item We show that FiRM has lower storage and complexity than Sigries, as it
    relies on simple SRAM filters instead of Misra-Gries. FiRM-P
    has zero slowdown for benign workloads and the same as
    Sigries under attack.
 
    \item We show that filtered Rowhammer mitigation need not rely on a
    probabilistic fallback. {\em FiRM-D} is a deterministic filtered Rowhammer
    mitigation that avoids reliance on RNG and has less than two-thirds the storage of Sigries.
\end{enumerate}
 
%\noindent
Our paper shows that neither the security tradeoff nor the CAM complexity of Sigries is essential for designing an effective filtered Rowhammer mitigation. A principled design can deliver both strong security and simplified implementation.

% ============================================================================
\clearpage

\section{Background and Motivation}
\label{sec:background}

% ---------------------------------------------------------------------------
\subsection{Threat Model}
\label{sec:threat}

We assume an attacker who knows the mitigation algorithm and all of its parameters (tracker size, thresholds, sampling probability, and the mode). The only secret is the output of the random number generator (RNG) used by a probabilistic mitigation. A design is
\emph{secure} if it can guarantee that the number of activations between consecutive mitigation/refresh remains no more than $T_{RHD}$.  RowPress~\cite{rowpress} is not considered, as it can be handled with page-closure policy~\cite{saxena2024impress}.

% ---------------------------------------------------------------------------
\subsection{DRAM Architecture and Parameters}
\label{sec:dram}

Accessing DRAM requires an ACT to move a row into the row-buffer, and two ACTs
to the same bank must be separated by the row-cycle time $t_{RC}$. To retain
data, every row must be refreshed within a refresh window $t_{REFW}$; the MC
issues a refresh every $t_{REFI}$, which occupies the bank for $t_{RFC}$.
Table~\ref{tab:dram} shows the DDR5 parameters. An attacker can do up to 620K activations to bank within the refresh window $t_{REFW}$.

\begin{table}[htb]
\centering
\vspace{-0.1 in}
\caption{DDR5 Parameters (DDR5-6000AN, 32-Gigabit chip)~\cite{micron_ddr5}}
\label{tab:dram}
\footnotesize
\begin{tabular}{l l r}
\toprule
{\bf Parameter} & {\bf Description} & {\bf Value} \\
\midrule
$t_{REFW}$   & Refresh window                         & 32\,ms \\
$t_{REFI}$   & Interval between REF commands          & 3900\,ns \\
$t_{RC}$     & Time between successive ACTs to a bank & 46\,ns \\
$t_{DRFMsb}$ & Same-bank DRFM latency (8 banks)       & 240\,ns \\
$t_{DRFMab}$ & All-bank DRFM latency (32 banks)       & 280\,ns \\
\bottomrule
\end{tabular}
\end{table}

% ---------------------------------------------------------------------------
\subsection{Rowhammer and Thresholds}
\label{sec:rh}

Rowhammer~\cite{kim2014flipping} occurs when frequent activations of an
\emph{aggressor} row cause bit-flips in nearby \emph{victim} rows. The Rowhammer
Threshold is typically quoted for single-sided ($T_{RHS}$) or
double-sided ($T_{RHD}$) patterns. While thresholds dropped from 140K ($T_{RHS}$) in
2014~\cite{kim2014flipping} to 4.8K ($T_{RHD}$) in 2020~\cite{kim2020revisitingRH},
 recent DDR5 characterization shows $T_{RHD}$ has stayed within 5K--12K range since
then~\cite{decadeRH,meyer2026phoenix}. The Sigries paper~\cite{sigries} observes that {\em the DRAM industry does not expect thresholds to fall below a few thousand}. 
Like Sigries, we therefore target thresholds of a few thousand
activations: our
{\bf default $T_{RHD}$ is 3K}, with sensitivity to 2K and 4K.

\subsection{Mitigation: DRFM and Aggressor Detection}
\label{sec:mitig}

To mitigate Rowhammer at the Memory-Controller (MC), we need to track aggressor rows and refresh the victim rows. 

\vspace{0.05 in}
\noindent{\bf{DRFM:}} The MC cannot refresh a victim row directly, as the
aggressor-to-victim mapping~\cite{DRAMSecrecy,DRAMA} is internal to the DRAM. DDR5 provides \emph{Directed Refresh Management} (DRFM)~\cite{dream,qazi2025drfm}, which lets the MC sample the given row (using a special precharge command) to the DRAM chip and then issue the DRFM command to refresh the victims associated with the sampled row. \emph{DRFMsb}
mitigates the row of one bank per eight bank-groups and stalls those
8 banks for 240\,ns, whereas \emph{DRFMab} mitigates and stalls all 32 banks of the
given rank for 280\,ns. As DRFM stalls multiple banks, issuing a DRFM to mitigate only a single bank is inefficient.  Therefore, DREAM~\cite{dream} recently proposed delaying the issue of DRFM until the bank needs to sample another row, thus allowing sibling banks more time to perform their own sampling (and later do concurrent mitigation under a single DRFM), resulting in reduced number of DRFM and associated stalls.

\vspace{0.05 in}
\noindent{\bf{PARA:}}  PARA~\cite{kim2014flipping} is a probabilistic method to identify aggressor rows. On an activation, the MC mitigates the
neighbors of the activated row with probability $p$. PARA has a non-zero probability of failure. 
We target a failure rate comparable to that of naturally occurring errors,
namely 1 failure per 10K years per bank~\cite{ddr4errors}. For this target,
we use $p = 20/T_{RHD}$, which closely matches the rate obtained by the
recurrence-based method of Saroiu and Wolman~\cite{Sampling}
(see Appendix~\ref{app:para-sample}).\footnote{We note that for a victim row to fail under the double-sided Rowhammer attack, both aggressors must escape mitigation during $T_{RHD}$ activations each, so this becomes equivalent to a single-sided attack with $p$ set to $40/T_{RHS}$.} 

Table~\ref{tab:para} shows the selection probability $(p)$ and the overhead of PARA for Naive-DRFM (issue DRFM after sampling) and DREAM (delay DRFM until next sampling) under a pattern that performs continuous activations. We assume DRFMsb, so, on average, we incur a delay of 240 ns after $1/p$ activations to the bank (which requires $46/p$ ns). At $T_{RHD}$ of 3K, Naive-DRFM causes a loss of 28\% (8x stalls), whereas DREAM-DRFM causes a loss of only 3.5\%. So, we assume PARA is always implemented with DREAM.

\begin{table}[htb]
\centering
\setlength{\tabcolsep}{6pt}
\vspace{-0.1 in}
\caption{PARA parameters and overhead.}
\label{tab:para}
\begin{tabular}{cccc}
\toprule
 &  & \multicolumn{2}{c}{BW Overhead} \\
\cmidrule(l){3-4}
T\textsubscript{RHD} & $p$ & Naive-DRFM & DREAM-DRFM \\
\midrule
2000 & $1/100$ & 41.7\% & 5.2\% \\
{\bf 3000} & {\bf $1/150$} & {\bf 27.8\%} & {\bf{3.5\%}} \\
4000 & $1/200$ & 20.9\% & 2.6\% \\
\bottomrule
\end{tabular}
\end{table}

\noindent{\bf{Misra-Gries (MG):}} PARA incurs slowdown even when the system is not under an attack.  The slowdown from mitigations can be reduced by using precise tracking.   The most prominent tracker is the Misra-Gries~\cite{MG} algorithm used in Graphene~\cite{park2020graphene}. MG tracker contains a table with $K$ tagged-entries (each with a counter) and a {\em spill-counter}. On activation, if the row is contained in the table, the corresponding counter is incremented.  Otherwise, the spill-counter is incremented, and if it is above the minimum-count entry in the table, that entry is overwritten with the accessed row and spill-counter. A mitigation is issued if the tracked count crosses a specific internal threshold ($T_{MG}$). As the tracker is periodically reset (e.g. every tREFW), this causes the tracker to forget the counts of the tracked row.  An attacker could do ($T_{MG}-1$) activations each before and after reset, and evade mitigation even after $2(T_{MG}-1)$ activations.  For safe reset, $T_{MG}$ is set to $T_{RHD}/2$.

\begin{table}[htb]
\centering
\setlength{\tabcolsep}{6pt}
\caption{Tracker threshold and storage for Misra-Gries.}
\label{tab:mg}
\begin{tabular}{ccccc}
\toprule
T\textsubscript{RHD} & T\textsubscript{MG} & Entries & Entry Size & Storage/Bank \\
\midrule
2000 & 1000 & 623 & 29 bits & 2258 Bytes \\
\textbf{3000} & \textbf{1500} & \textbf{416} & \textbf{30 bits} & \textbf{1560 Bytes} \\
4000 & 2000 & 312 & 30 bits & 1170 Bytes \\
\bottomrule
\end{tabular}
\end{table}

Table~\ref{tab:mg} shows the $T_{MG}$, the number of table entries, and the size per bank for the Misra-Gries tracker as $T_{RHD}$ is varied from 2K to 4K. We note that each entry of MG table contains a valid bit, lock bit (to lock rows that have crossed $T_{MG}$), a 17-bit tag, in addition to the counter.  The problem with MG tracking is the large number of entries per bank (416 for our default $T_{RHD}$ of 3K), which must be searched 
associatively on {\em every} activation. The CAM (content-addressable memory) structure required to design the MG table results in significant area, complexity, and power consumption. This is the main reason unrestricted MG tracking has not been adopted commercially, even though it is a principled solution that issues only a minimal number of mitigations.

% ---------------------------------------------------------------------------
\subsection{Sigries: Rowhammer Mitigation in Production}
\label{sec:sigries}

PARA suffers from frequent mitigations, whereas MG suffers from the complexity of large CAMs.  A recent paper~\cite{sigries} outlines six requirements for an MC-based Rowhammer mitigation to be viable for commercial adoption:  

\vspace{0.05 in}
\noindent{\bf Requirements:} \textbf{(R1)} Minimal bandwidth and latency overheads when not under attack, \textbf{(R2)} No performance outliers, \textbf{(R3)} Liveness under attack, no microsecond-scale delays, \textbf{(R4)} Low hardware cost, \textbf{(R5)} Flexibility to handle different hardware via simple configuration changes, and \textbf{(R6)} Configurable security guarantees that bound the exposure to a desired level.

\begin{figure}[!htb]
    \centering
    \vspace{-0.05 in}
\includegraphics[width=3.25in]{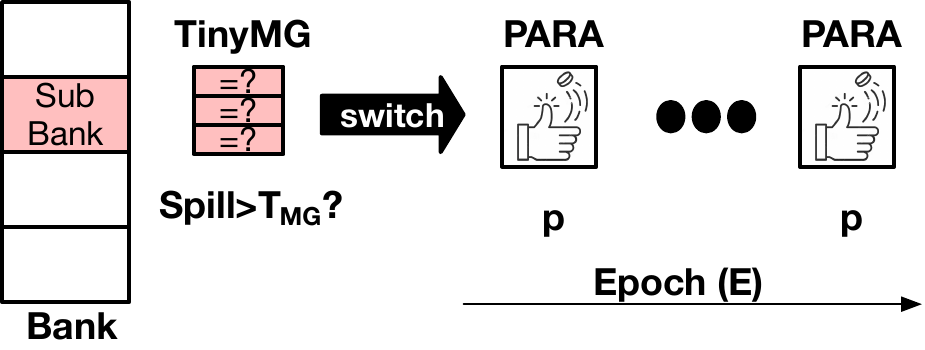}
    \caption{Overview of Sigries: It divides the bank into several sub-banks, each provisioned with TinyMG. If the spill-counter of TinyMG exceeds $T_{MG}$, it switches to PARA and remains there for at least E windows, exiting only if
its shadow counters indicate the attack has stopped.}
    \label{fig:sigries}
\end{figure}

\vspace{0.05 in}
\noindent{\bf Design:} Sigries~\cite{sigries}, deployed by Microsoft in Azure Cobalt
200, is an MC-side defense, developed against
these six requirements. Figure~\ref{fig:sigries} shows the overview of Sigries. Sigries divides each bank into $S$ sub-banks, each equipped with an under-provisioned MG tracker (TinyMG) containing only a couple dozen or so entries. TinyMG is secure as long as the spill-counter remains below $T_{MG}$. We call this {\em lite-mode} as it incurs minimal mitigation overheads.  However, when the spill-counter reaches $T_{MG}$, Sigries transitions to a mode where each activation gets mitigated with probability $p$. This mode is called {\em heavy-mode} as it can incur frequent mitigation. Sigries remains in heavy-mode for an Epoch of at least $E$ windows. While in heavy-mode, Sigries uses {\em shadow counters} to track the intensity of the attack on the sub-bank, and extends the epoch beyond $E$ if the attack appears to be ongoing.

Each mode is designed to be secure {\em in isolation},
assuming a row enters the mode with zero unmitigated activations. Therefore, $T_{MG}$ is set to $T_{RHD}/2$ and the p is set to 20/$T_{RHD}$.

\vspace{0.05 in}
\noindent{\bf Security:}
Sigries
recognizes that the lite-mode to heavy-mode transition can cause insecurity, as a row may start with $T_{MG}$ activations already consumed, which may not leave enough activations for PARA to operate securely. 
Sigries treats this as a bounded \emph{vulnerability window} of one tREFW and amortizes it by using the heavy-mode for at least $E$ windows after each transition (Epoch length gets extended under continued attacks). Thus, the vulnerability can be bounded to at most 1/E.  Sigries
picks $E$ such that the vulnerability is bounded per sub-bank (granularity was confirmed to us by its authors) to be within 1 hour per year (E $\ge$ 8760 windows).

\vspace{0.05 in}
\noindent{\bf Performance:}  The key advantage of Sigries is that it provides zero slowdown for benign applications, and the mitigation overhead remains small even under attack.

\vspace{0.05 in}
\noindent{\bf Storage and Complexity:}  As the Sigries paper does not disclose the parameters of the deployed
design, all Sigries parameters used in this paper (including $T_{MG}$,
$p$, $E$, and the tracker size) are {\bf estimated or inferred} from its design
description for our target $T_{RHD}$ of 3K. Without loss of generality, we assume that each TinyMG is configured to use 32 entries (which is compatible with {\em couple/few-dozen entries} mentioned in the Sigries paper). We then vary the number of sub-banks and identify those for which there is never a switch from lite-mode to heavy-mode,
which for our system means S $\ge$ 8. Table~\ref{tab:sigentries} shows the number of entries in Graphene and Sigries (with 8 and 16 sub-banks).  We use Sigries with 8 sub-banks and equip each sub-bank with a 32-entry TinyMG.

\begin{table}[htb]
\centering
\setlength{\tabcolsep}{6pt}
\caption{Number of MG Entries in Graphene and Sigries ($T_{RHD}$ of 3K).}
\label{tab:sigentries}
\begin{tabular}{cccc}
\toprule
Design & Sub-banks & Entries/Sub-bank & Total Entries/Bank \\
\midrule
Graphene & 1 & 416 & 416 \\
Sigries & 8 (default) & 32 & 256 \\
Sigries & 16 & 32 & 512 \\
\bottomrule
\end{tabular}
\end{table}

\subsection{Goal of This Paper}
\label{sec:goal}

Sigries is the first production-level Rowhammer defense described in detail in a research paper, which allows the scientific community to review the industrial design. To that end, the goal of our paper is to answer the following questions:

\vspace{0.05in}

\noindent{\bf (1)} Does the Sigries design have any additional unknown vulnerabilities? This is important as Sigries is not a theoretical design
but something deployed in commercial systems.
\vspace{0.05in}

\noindent{\bf (2)} What is the impact of Sigries vulnerability target (of one
hour per year) on the Mean-Time-To-Failure of the system?
\vspace{0.05in}

\noindent{\bf (3)} Can we design an MC-side mitigation that provides strong security (no vulnerability windows), avoids the CAM complexity of Sigries, and meets the
performance of Sigries?
\vspace{0.05in}

\noindent{\bf (4)} Can such a dual mode scheme be made fully deterministic, thus
avoiding the reliance on any probabilistic fall-back?

\ignore{
The goal of our paper is to answer these questions.  We also aim to deveo, and  Our goal is a principled filtered mitigation in which the
filter and the fallback are {\em co-designed}, so that the activation budget of
an attacker is bounded not only within each mode but also across both
transitions, leaving no vulnerability window. We also seek to remove the CAM
entirely, replacing exact heavy-hitter tracking with an untagged, direct-mapped
SRAM filter that merely tests whether traffic to a region is below a filtering
threshold $T_F$, while retaining the defining property of a filtered design:
zero mitigations, and hence zero slowdown, for benign workloads.
\S\ref{sec:vuln} answers questions (1) and (2), and the rest of the paper
answers (3) and (4).

}
\newpage
\section{Analyzing Security of Sigries}
\label{sec:attack}

%Furthermore, the authors identified one of the transitions (from lite-mode to heavy-mode) as a source of vulnerability. 

The two modes of Sigries are designed to be secure individually.
The Sigries paper mentions
that the transition from heavy-mode to lite-mode is {\em always} safe. In
this section, we show a new vulnerability that targets the other transition. Furthermore, we also analyze
the impact of the default vulnerability target of Sigries on the system Mean-Time-To-Failure with an attack that targets multiple sub-banks.

\subsection{Attacking the Second Transition}

\vspace{0.05 in}
\noindent{\bf The budget of lite-mode.} On entering lite-mode, the TinyMG tracker
is reset. A row can therefore accrue at most $T_{MG}$ activations before
the tracker forces a mitigation. As Sigries does not disclose $T_{MG}$, we credit it with $T_{MG} = T_{RHD}/2$, to securely handle the reset of the tracker against an attacker who
straddles a reset with $T_{MG}$ activations on either side. Lite-mode is thus secure {\em exactly} at its design
point, with no slack, and only if the row enters with zero
unmitigated activations.

\vspace{0.05 in}

\noindent{\bf The budget of heavy-mode.} PARA is configured with
$p = 20/T_{RHD}$. This probability is chosen so that a row is very
unlikely to reach $T_{RHD}$ activations without a mitigation. Critically,
PARA is provisioned against a budget of $T_{RHD}$, not $T_{MG}$: by its
own design point, a row in heavy-mode is permitted to accumulate up to
$T_{RHD}$ unmitigated activations, which is $2\times$ the budget that
lite-mode can absorb.

\vspace{0.05 in}

\noindent{\bf The composition is unsafe.} Let $C$ denote the unmitigated
activations a row carries out of heavy-mode. The total activations the row
can accrue between two refreshes of the victim is $C + T_{MG}$. Security
requires $C + T_{MG} \leq T_{RHD}$, i.e. $C \leq T_{MG}$. But nothing in
heavy-mode enforces this: PARA is calibrated to bound $C$ at $T_{RHD}$,
which is twice the permissible value for a safe entry into the lite-mode. At the design point of PARA, the total
reaches $1.5 \times T_{RHD}$. Sigries
deems this transition safe because it reasons about lite-mode in
isolation, assuming lite-mode starts with zero unmitigated activations, however, the guarantee is voided by what the
preceding mode hands it.

\begin{figure}[!htb]
    \centering
\includegraphics[width=3.25in]{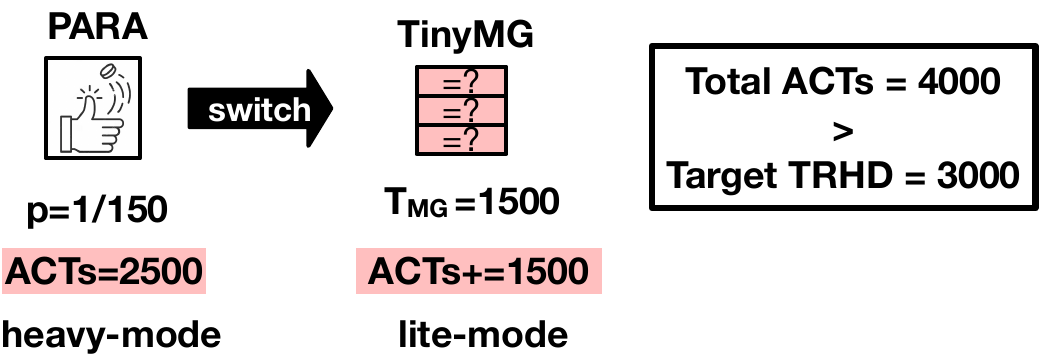}
    \caption{Example attack: In heavy-mode, PARA permits 2500 unmitigated activations (within TRHD of 3000), then the lite-mode permits an additional 1500 activations, for a total of 4000 activations, which violates security.   }
    \label{fig:attack}
\end{figure}

\noindent{\bf The attack.} Figure~\ref{fig:attack} shows our attack for the
default $T_{RHD}$ of 3K. PARA permits 2500 unmitigated activations on the
attacked row in heavy-mode, within the target of 3000. On switching back to
lite-mode, the row accrues an additional $T_{MG}$ = 1500 activations, for a
total of 4000 across the two modes without a single mitigation, which violates security.

%Figure~\ref{fig:attack} shows the overview of our attack.  For our default $T_{RHD}$ of 3K, $T_{MG}$ is 1500, and PARA is designed (with p=1/150) so as to permit no more than 3K unmitigated activations in the heavy-mode. The attacker hammers a target row in heavy-mode, and PARA permits 2500 unmitigated activations on the attacked row (which is still within the target TRHD).  However, when the design switches back to lite-mode, the row accrues an additional $T_{MG}$ (1500) activations, for a total of 4000 unmitigated activations across the two modes, without triggering any mitigation, which violates security. 

\vspace{0.05 in}

\ignore{
\noindent{\bf The required probability.} For the carry-in to be safe with the
budget that lite-mode leaves, PARA must bound $C$ at $T_{MG}$
rather than $T_{RHD}$, which requires $p \geq 20/T_{MG} = 40/T_{RHD}$. This is the first indication that to ensure security, the two modes should not be parameterized independently.
}

%The probability used by Sigries is therefore $2\times$ too small. 

\subsection{Understanding the Impact of the New Vulnerability}

As $p$ is calibrated with an escape 
probability of $e^{-20}$ per aggressor, the target victim failure probability under double-sided attack is $e^{-40}$. Under our attack, each aggressor carries a $T_{MG}$ budget across the
transition, so each aggressor fails with a probability of 
$(1-p)^{(T_{RHD} - T_{MG})} = e^{-10}$, and  the victim will fail with probability $e^{-20}$, which is $e^{20} = 4.9 \times 10^{8}$ higher than the steady-state failure rate of heavy-mode.

\ignore{
Notably, the disclosed lite-to-heavy transition carries {\em the same}
exposure. There, a row enters heavy-mode with $T_{MG}$ activations already
consumed, leaving $T_{RHD} - T_{MG} = T_{MG}$ activations for PARA to
cover, which it misses with probability $(1-p)^{T_{MG}} = e^{-10}$. The
two transitions are symmetric, as summarized in Table~\ref{tab:transition};
Sigries budgets for one and declares the other always safe. The true
exposure is therefore twice what is specified.

\begin{table}[h]
\centering
\setlength{\tabcolsep}{6pt}
\caption{Exposure of the two mode transitions ($T_{RHD}$ of 3K).}
\label{tab:transition}
\begin{tabular}{lccc}
\toprule
Transition & Budget Left & Failure Prob. & Disclosed \\
\midrule
Lite $\rightarrow$ Heavy & $T_{MG}$ & $4.5\times10^{-5}$ & Yes \\
Heavy $\rightarrow$ Lite & $T_{MG}$ & $4.5\times10^{-5}$ & No \\
\bottomrule
\end{tabular}
\end{table}
}

As the vulnerability exists at both transitions, an epoch of $E$ windows
contains two vulnerable windows rather than one, doubling the exposure of
Sigries to two hours per year per sub-bank. More importantly, a budget
expressed as a few hours per year {\em per sub-bank} is an inadequate way
to assess system security, as it bounds how long one sub-bank is exposed
and says nothing about the failure rate of a system with thousands of
them. The next section shows that such a target
results in an unacceptably low Mean-Time-To-Failure (MTTF) under a {\em
Round-Robin Attack} that attacks all sub-banks in turn.

\subsection{Round-Robin Attack: A Window is Open Most of the Time}

%The vulnerability target of Sigries is {\em per sub-bank}. At the default
%$E = 8760$, each sub-bank is exposed for one $t_{REFW}$ out of every
%$E$, or one hour per year. This bound looks sound in isolation, but an
%attacker can target many sub-banks.

\begin{figure}[!htb]
    \centering
        \vspace{-0.05 in}

\includegraphics[width=3.25in]{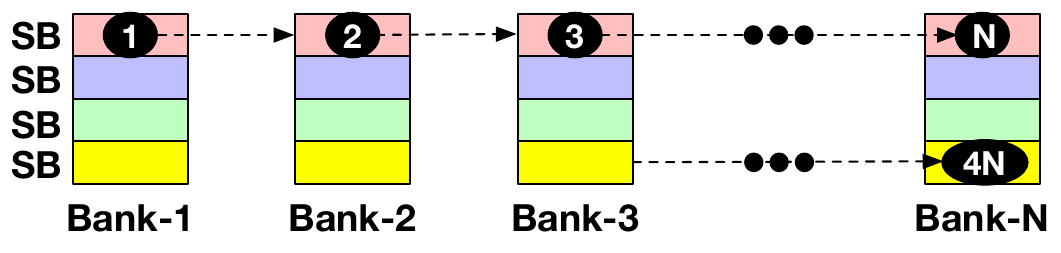}
    \vspace{-0.05 in}

    \caption{Round-Robin Attack on a system with N banks, and 4 sub-banks per bank.  It attacks only one sub-bank per tREFW, goes round-robin across all sub-banks, and goes through all the sub-banks in 4N tREFW.}
    \vspace{-0.05 in}
    \label{fig:roundrobin}
\end{figure}

The default Epoch size (without continuous attack on the sub-bank) is 8760.  Ideally, if there were 8760 sub-banks in the system, we could devise a pattern that goes through all the sub-banks in a {\em Round-Robin} fashion, attacking only one sub-bank at a time, as shown in Figure~\ref{fig:roundrobin}. We observe that Cobalt 200 has 12 memory channels~\cite{cobalt200}, each DDR5 channel has 64 banks, and with 8 sub-banks per bank, we have 12x64x8=6144 sub-banks, slightly less than the 8760 we need.

%\footnote{We pessimistically assume that Sigries has 8 sub-banks per bank. If the actual Sigries design in Cobalt 200 had 16+ sub-banks per bank, then we would meet the requirement and the attack would obtain a vulnerability window open at every tREFW.}

Our {\em Round-Robin Attack}
 targets the 6144 sub-banks, staggered by one $t_{REFW}$ each, so in any $t_{REFW}$  only one sub-bank is under attack. It then waits
the remaining $8760 - 6144 = 2616$ windows for the first sub-bank to fall
back to lite-mode, and repeats. A vulnerability window is thus open
$6144/8760 = {\bf{70}}\%$ {\bf{of the time}}, rather than the $1/8760$ that the
per-sub-bank budget suggests. To attack a sub-bank within $t_{REFW}$ we perform 4K activations to each of the 33 rows that map to the same sub-bank in a circular fashion.

\vspace{0.05 in}
\noindent{\bf Evading the Attack Detector:} Sigries extends the epoch when its shadow counters indicate that an attack is still ongoing. Our pattern bypasses this detector.  A sub-bank gets hammered in one window and receives zero activations for next $E-1$ windows.  As only one window per epoch appears anomalous, each sub-bank would return to lite-mode after $E$ windows.\footnote{The success of our attack does not hinge on the exact Epoch size. Even if every entry into heavy-mode elongated the epoch by $10\times$, the MTTF under our attack would be 10 seconds, still seven orders of magnitude below PARA.}

%Our attack is specifically designed to evade the detection mechanism of Sigries as it performs zero activations when the sub-bank is not under attack. 

%%%%%%%%%%%%%%%

\subsection{Analyzing the MTTF Under Round-Robin Attack}
\label{sec:mttf}
 
\begin{figure}[!htb]
    \centering
    \vspace{-0.1 in}
\includegraphics[width=2.5in]{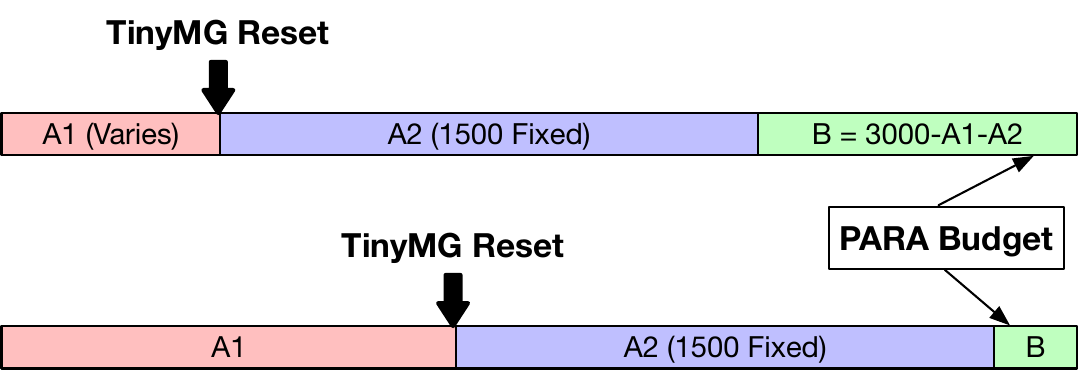}
    \caption{Activation budget (TRHD) is split into three parts: A1 (before TinyMG reset), A2 to overflow TinyMG, and B is the leftover budget for PARA.  In windows where A1 is high, B is severely low for PARA-based sampling.}
    \label{fig:split}
\end{figure}

The key challenge in determining the MTTF of Sigries is converting a vulnerability
window into the probability that a victim row fails. Assuming the sub-bank is already in the lite-mode, when the attack starts, the total activation to each row gets distributed into three parts: (1) The activations ($A_1$) that may occur before the reset of the TinyMG, (2) The activations ($A_2$) that are needed for overflowing the TinyMG tracker, and (3) The leftover budget (B) for PARA to perform sampling-based mitigation.  Figure~\ref{fig:split} shows such a split.

We pessimistically assume\footnote{An attacker can learn the periodic schedule of TinyMG reset using side channels, and synchronize~\cite{meyer2026phoenix} the attack such that $T_{MG}-1$ activations to each aggressor are performed right before the reset of the TinyMG tracker, thereby exhausting almost all the activation budget and leaving almost nothing for PARA.  This would increase the probability of failure for each vulnerability window to almost 1.  We do not consider such an attack. } that the attack is not synchronized with the reset of the TinyMG tracker (such reset occurs periodically, once per $t_{REFW}$), so for some of the attack windows, the reset of the tracker will cause a non-zero ($A_1$) number of activations to be forgotten from the tracker.  By design, $A_2$ equals $T_{MG} = T_{RHD}/2$.  Thus, the leftover activation budget for PARA,
$B = T_{RHD} - A_1 - A_2$. We note that the activations for $A_1$ and $A_2$
escape mitigation by construction. Only $B$ is sampled, so the failure
probability of the victim is decided by the budget available to PARA.

\vspace{0.05 in}

\noindent{\bf The share of $A_1$ is not fixed:} We note that as our attack is not synchronized with the TinyMG reset, $A_1$ is a dynamic quantity. 
The tracker reset is periodic in $t_{REFW}$ and is orchestrated by the memory controller, without regard to the access pattern. The attack is therefore not synchronized with the reset, which may land anywhere within it. As
the pattern delivers activations at a constant rate, and the tracker
forces a mitigation once a row reaches $T_{MG}$, if a reset happens before the overflow of the TinyMG tracker, the value of $A_1$ would be uniformly
distributed on $[0, T_{MG}]$, and so the budget left to sampling is also uniformly distributed,
$B = T_{MG} - [0, T_{MG}]$.

\vspace{0.05 in}

\noindent{\bf Count chances for PARA, not activations.} It is intuitive to think of PARA in terms of the number of chances for mitigation that the budget provides. A budget of $B$ activations buys the defense, on average, $\lambda = p \cdot B$ {\em chances} to catch the aggressor, and the aggressor escapes only if all of them miss. At $p=1/150$, a budget of 1500 provides 10 chances and the likelihood of missing all is $e^{-10}$, whereas a budget of 150 provides one chance and the likelihood is $e^{-1}$. Reducing the budget causes exponential decay in the effectiveness of PARA.

\begin{table}[htb]
\centering
\caption{Escape probability as the reset donates more of the budget. Each
decile occurs one time in ten; $B$ and $\lambda$ are at its midpoint.}
\label{tab:delta}
\footnotesize
\begin{tabular}{ccccc}
\toprule
$A_1$ (\% of $T_{MG}$) & $B$ & $\lambda = p\,B$ & ProbAggRow & ProbVicFail \\
\midrule
0--10 & 1425 & 9.5 & $7.3\!\times\!10^{-5}$ & $5.3\!\times\!10^{-9}$ \\
10--20 & 1275 & 8.5 & $2.0\!\times\!10^{-4}$ & $3.9\!\times\!10^{-8}$ \\
20--30 & 1125 & 7.5 & $5.4\!\times\!10^{-4}$ & $2.9\!\times\!10^{-7}$ \\
30--40 & 975 & 6.5 & $1.5\!\times\!10^{-3}$ & $2.2\!\times\!10^{-6}$ \\
40--50 & 825 & 5.5 & $4.0\!\times\!10^{-3}$ & $1.6\!\times\!10^{-5}$ \\
50--60 & 675 & 4.5 & 0.011 & $1.2\!\times\!10^{-4}$ \\
60--70 & 525 & 3.5 & 0.030 & $8.9\!\times\!10^{-4}$ \\
70--80 & 375 & 2.5 & 0.081 & $6.6\!\times\!10^{-3}$ \\
80--90 & 225 & 1.5 & 0.222 & 0.049 \\
90--100 & 75 & 0.5 & 0.606 & 0.367 \\
\midrule
\textbf{Average} & --- & --- & \textbf{0.096} & \textbf{0.042} \\
\bottomrule
\vspace{-0.2 in}
\end{tabular}
\end{table}

Table~\ref{tab:delta} varies $A_1$ across its
range in steps of one tenth; as $A_1$ is uniform, each row occurs one time
in ten, so their average is the escape probability the attacker faces. The last two columns give the resulting failure probabilities: \texttt{ProbAggRow} $=(1-p)^{B}\approx e^{-\lambda}$ is the chance one aggressor escapes all $\lambda$ chances, and, as a double-sided attack needs {\em both} aggressors to escape, \texttt{ProbVicFail} $=$ \texttt{ProbAggRow}$^2$. For the first five rows the escape probability is negligible; the risk is concentrated in the last two, where fewer than one expected mitigation remains and the sampler almost always misses. The final row alone contributes $0.061$ of the $0.096$ average. As each vulnerability window draws a different $A_1$, averaging \texttt{ProbVicFail} over its uniform distribution gives a victim-row failure probability of 0.042 per vulnerability window.

\vspace{0.05 in}
\noindent{\bf The Resulting MTTF.} If a victim fails in a vulnerability window with 4.2\% probability, then we need 24 vulnerability windows on average to cause a failure.  Given 70\% of the windows are vulnerable, we need 24/0.7 = 34 windows.  As each $t_{REFW}$ window is 32 milliseconds, {\bf the Mean-Time-to-Failure is about 1 second}. For comparison, we configured PARA with $p$ based on {\em one
failure per 10K years per bank}, which for Cobalt 200 means a
system MTTF of 13 years under continuous attack on all banks.\fix{\footnote{The comparison understates the gap: the 13-year MTTF for PARA assumes all 768 banks are activated continuously at maximum rate (100\% of system activation budget), whereas our attack touches one bank per $t_{REFW}$ at one-fifth of its maximum rate, consuming under 0.03\% of system activation budget.}} Thus, relative to PARA,
Sigries {\bf lowers the MTTF by eight orders of magnitude}.
 
\subsection{Takeaway: Duty Cycle Bounds Exposure, Not Failure}
\label{sec:takeaway}

The security of Sigries rests on $E \geq 8760$, which bounds {\em how often} a vulnerability window is open but says nothing about the probability of failure {\em within} it, which we show to be $4.2\%$ rather than the $e^{-40}$ that heavy-mode targets. Once the Round-Robin Attack raises the fraction of vulnerable windows to $70\%$, the in-window failure probability results in an unacceptably low guarantee in practice.

\vspace{0.05 in}

Ideally, we want an MC-side mitigation that avoids both the vulnerability windows of Sigries and the CAM complexity of Sigries. The next section develops such a solution.

\clearpage

\ignore{
Prolog: Why and the two insights
Design and Operation: Structures and how it works?
Parameters: APM and ATH
Security Analysis: Two modes and two transition
Storage Analysis:
Experimental Method:
Results (Benign):
Results (Attack):
}

\section{FiRM: Filtered Rowhammer Mitigation}
\label{sec:firm}

In this section, we design a fully secure and complexity-effective Rowhammer mitigation.  Our proposal, {\em FiRM (Filtered Rowhammer Mitigation)},  is based on two key insights:

\vspace{0.05 in}
\noindent{\bf Insight-1: The two modes share one budget.} For security, the quantity that
must be bounded is the number of activations a row receives between two
mitigations, and this budget is consumed by whichever mode
happens to be active. Sigries allocates the full budget $T_{RHD}$ to each
mode separately, so a row that visits both can spend it twice. FiRM instead
{\em partitions} $T_{RHD}$ across the {\em filter} and the {\em fallback}, so that the
sum over any mode sequence remains within the threshold.

\vspace{0.05 in}

\noindent{\bf Insight-2: The filter needs only one bit of information.} The TinyMG of Sigries is designed to
identify {\em which} row is hot, which is why it needs tags and an
associative search. But once the fallback carries the security guarantee, the
first stage only needs to determine if the fallback should be used or not. FiRM therefore
replaces the CAM-based filter of Sigries with an untagged, direct-mapped table of counters that tracks if the entry has accumulated a given number of activations. Crucially,
aliasing in such a table is {\em safe}: several rows sharing a counter cause
it to over-count, so the counter is an upper bound on the activations of any
row that maps to it. Over-counting can cost performance, but does not cost security. We note that, even for Sigries, the job of mode-switching is mainly regulated by the spill-counter (with $ c$ entries, the mode can switch after $(c+1)\cdot T_{MG}$ activations to the sub-bank).

\vspace{0.05 in}

We develop two variants of FiRM based on the type of fallback algorithm: {\em{FiRM-P (probabilistic)}} and {\em{FiRM-D (deterministic)}}.  In this section, we focus on FiRM-P and the next section discusses FiRM-D.

\subsection{FiRM-P:  Design and Operation}

\begin{figure}[!htb]
    \centering
\includegraphics[width=3.5in]{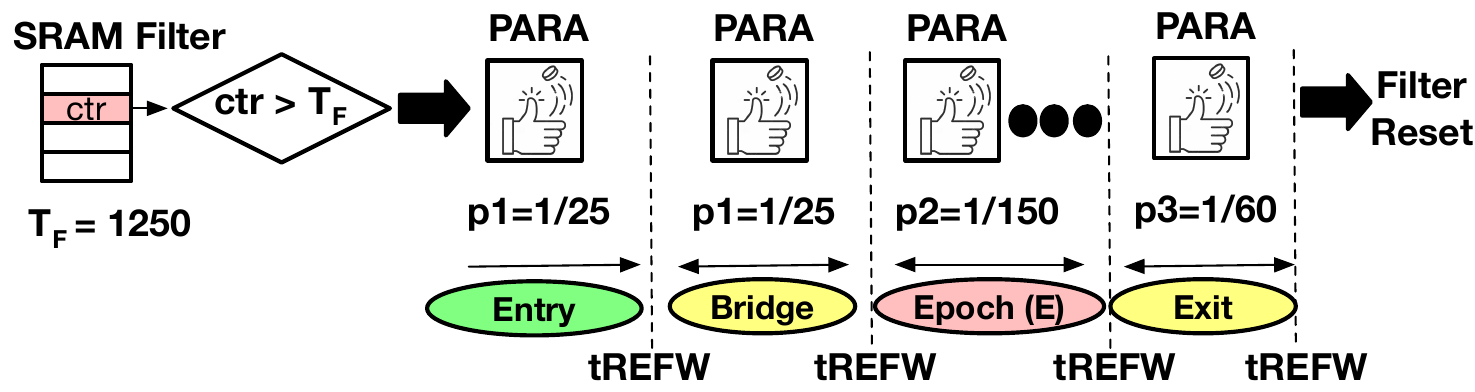}
    \caption{Overview of FiRM-P. FiRM-P splits the activation budget (3K) between the filtering threshold ($T_F$) and PARA. On crossing $T_F$ it samples at $p_1$=1/25 for the rest of the {\em entry} window and for one further {\em bridge} window (with the filter disabled), then spends $E$ windows at $p_2$=1/150, and one {\em exit} window at $p_3$=1/60 (based on $T_F$) before resetting the filter and reusing it.}
    \label{fig:firmp}
    \vspace{-0.1 in}
\end{figure}

Figure~\ref{fig:firmp} shows an overview of FiRM-P. FiRM uses an SRAM-based filter containing $F$ entries. This is equivalent to a bank containing $F$ sub-banks, each with one filter entry.  The filter entry is simply an activation counter that is incremented on activation to the region. Thus, our filter avoids tags and the complexity of CAM. If the value of the accessed counter is below a {\em Filtering Threshold ($T_F$)}, no mitigation is needed. This is the {\em lite-mode}. However, if the count exceeds $T_F$, FiRM-P enters {\em heavy-mode} where probabilistic mitigation is employed. The design remains in {\em heavy-mode} for the rest of that window and for $E+2$ full windows thereafter. Note that the value of E determines the asymptotic performance under attacks and does not impact security.

\vspace{0.05 in}
\noindent{\bf Filtering Threshold.} $T_F$ is bounded from above by
lite-mode security and from below by benign behavior. As the filter is
cleared every $t_{REFW}$ and the refresh may not be aligned with filter-reset,  an attacker can accrue $T_F$ activations on either side of the reset,
so lite-mode alone requires $2 \cdot T_F \leq T_{RHD}$. From below, $T_F$
must exceed the largest counter value any benign workload produces,
otherwise the workload trips into heavy-mode and pays mitigations it does
not need. For our default $T_{RHD}$ of 3K, we use $T_F = 1250$, which
satisfies the security bound of 1500 with a margin of 500 activations left for the heavy-mode within the given tREFW window.

\vspace{0.05 in}

\noindent{\bf Non-uniform sampling.} Unlike Sigries, FiRM-P does not use a
single $p$ throughout heavy-mode. It instead relies on three distinct
probabilities: $p_1$ (entry into heavy-mode), $p_2$ (steady-state in
heavy-mode), and $p_3$ (exit to lite-mode). A row can enter heavy-mode
carrying up to $2 \cdot T_F$ activations accrued because of the filter
(doubled due to filter reset), so the budget available to sampling is
reduced within that tREFW window. So, FiRM-P uses a reduced value of $p_1$
(e.g., $p_1$=1/25) until the tREFW window finishes. As the filter can
cross $T_F$ in the middle of a window, a row can carry up to $T_F$
activations into the {\em next} window as well, so FiRM-P spends one more
full window, which we call the {\em bridge} window, at $p_1$ with the
filter disabled ($T_F$=0). In subsequent steady-state (E) windows, PARA
has the full activation budget, so FiRM-P uses a low value of $p_2$ (e.g.,
$p_2$=1/150). Finally, to ensure that the heavy-mode to lite-mode
transition remains safe, there is one additional {\em exit} window executed with probability $p_3$, which is determined by $T_F$ instead of
$T_{RHD}$ (e.g. $p_3$=1/60).

\subsection{Impact of Epoch on Effective Probability of Heavy-Mode}

In heavy-mode, Sigries samples at $p = 1/150$ (for $T_{RHD}$ of 3K), and
this rate determines its slowdown. FiRM-P samples at $6\times$ this rate
during the entry and bridge windows, so the slowdown in those
windows is correspondingly higher. However, a transition window is followed by a long
run of steady-state windows sampled at $p_2$, which is identical to the rate
used by Sigries. The elevated rate is therefore amortized, and the average
mitigation rate of FiRM-P is similar to Sigries.

Table~\ref{tab:epoch} shows the effective probability of both designs
as $E$ is varied from 100 to 10K. At $E = 100$ FiRM-P issues 11\% more
mitigations than Sigries, but the gap closes to within $1\%$ by $E = 1000$. At the operating point of Sigries
($E \ge 8760$), the two designs are indistinguishable in mitigation rate.

\begin{table}[htb]
\centering
\vspace{-0.1 in}
\setlength{\tabcolsep}{6pt}
\caption{Average mitigation probability of FiRM-P and Sigries.}
\label{tab:epoch}
\begin{tabular}{cccc}
\toprule
Epoch ($E$) & FiRM-P ($p$) & Sigries ($p$) & Ratio \\
\midrule
%10    & $1/97$  & $1/150$ & 1.54$\times$ \\
100   & $1/135$ & $1/150$ & 1.11$\times$ \\
1000  & $1/148$ & $1/150$ & 1.01$\times$ \\
10000 & $1/150$ & $1/150$ & 1.00$\times$ \\
\bottomrule
\end{tabular}
\end{table}

The role of $E$ in the two designs is fundamentally different. For Sigries, $E$ is a {\em security}
parameter, which determines vulnerability. For
FiRM-P, $E$ is purely a {\em performance} parameter. Security is established
within every window, in both modes and across both transitions, by the
partition of the activation budget between the filter and the fallback. The next
section establishes the security guarantee formally.

%%%%%%%%%%%%%%%%%%%%%%%%%%%%%%%%%%%%%%%%%%%%%%%%%%%%

\subsection{Security Analysis}

We must show that no row receives more than $T_{RHD}$ activations
between two refreshes of the victim. Two facts simplify the argument.
First, a
refresh interval spans at most one reset and at most one mode boundary.
A row that sits just below $T_F$ when the filter is cleared starts
again from zero, so a single refresh interval can be charged {\em two}
filter allowances, one on each side of the clear.
Second, a fresh allowance of $T_F$ is granted only by a lite-mode window;
the bridge and heavy-mode windows run with $T_F = 0$. An
interval in which the filter contributes $A$ activations thus leaves
$T_{RHD} - A$ to the fallback, and is secure if the rate in force satisfies
$p \geq 20/(T_{RHD} - A)$. We apply this test to each mode and boundary of
Figure~\ref{fig:firmp}.

\vspace{0.05 in}

\noindent{\bf Case 1: Lite-mode.} A counter is an upper bound on the
activations of every row that maps to it (aliasing only makes it
over-count), so a row must be within $T_F$ activations after the filter
reset to stay in lite-mode. A refresh interval that straddles a
reset admits $T_F$ on either side, so $A = 2 \cdot T_F = 2500$,
which is within $T_{RHD}$. Security in lite-mode is thus {\em
deterministic} and does not rely on fallback.

\vspace{0.05 in}

\noindent{\bf Case 2: Heavy-mode.} Filtering is disabled, so $A=0$ and
the full budget is available to sampling: the rate may be as low as
$20/T_{RHD} = 1/150$, which is the steady-state rate $p_2$, which is sufficient to ensure security. The bridge and
exit windows sample faster, and the boundaries between heavy-mode windows
charge no allowance, so this case covers them.

\vspace{0.05 in}

\noindent{\bf Case 3: Entry (Lite-mode to Bridge).} The refresh interval
containing the crossing has a lite-mode prefix that may straddle
a filter reset, so $A = 2 \cdot T_F$, leaving only
$T_{RHD} - 2 \cdot T_F = 500$ activations for the suffix. Sampling at
$p_1 = 20/(T_{RHD} - 2 \cdot T_F) = 1/25$ bounds the suffix to this reduced
budget, and provides a failure rate that is equal to the target of PARA.
The filter and the fallback are thus charged 2500 and 500 activations,
which sum to exactly $T_{RHD}$.

\vspace{0.05 in}

\noindent{\bf Case 4: Bridge to Steady-State.} The entry window alone
is not sufficient to ensure security of steady-state heavy-mode, as the filter crossing can occur at any point within a window.
If it occurs near the end, the row carries its $T_F$ filter allowance into
the {\em next} window having paid almost no sampling for it. The interval
straddling that boundary is therefore charged $A = T_F = 1250$ with no new
allowance in its suffix, which requires $p \geq 20/1750 = 1/87$. The
bridge window supplies $p_1 = 1/25$ and is therefore secure. Relaxing to
$p_2$ at this boundary would not be secure: the suffix would fail with probability
$(1-p_2)^{1750} = e^{-11.7}$, more than three orders of magnitude above the
target. This is why the entry spends one full window at $p_1$ with
$T_F = 0$ before relaxing to $p_2$.

\vspace{0.05 in}

\noindent{\bf Case 5: Exit (Heavy-Mode to Lite-Mode).} The filter is reset on
exit, so the lite-mode suffix of the straddling interval contains no
further reset and contributes at most $A = T_F$. The heavy-mode prefix must
therefore be bounded by $T_F$ as well, which is what
$p_3 = 1/60 \ge 20/T_F$ provides, since
$(1-p_3)^{T_F} = e^{-21} < e^{-20}$. The exit is thus the mirror of
Case~1, with the carry-out of heavy-mode playing the role of the first
$T_F$ and the lite-mode suffix the second, for the same total of
$2 \cdot T_F$. As the epoch counter is known to the controller, the exit
window is identified in advance and $p_3$ applied to it.

\vspace{0.05 in}

\noindent{\bf No vulnerability windows.} In every case the activation
budget is partitioned so that the filter and the fallback together stay
within $T_{RHD}$, and the resulting failure probability never exceeds the
target rate of $e^{-20}$. FiRM-P therefore has no window in which the
guarantee is weaker than in steady state, and needs no vulnerability duty-cycle argument
to bound its exposure. Consequently, the multi-sub-bank attack of Sigries does not apply: there is no window for an
attacker to schedule. This
also explains why $E$ is free of security implications, as it does not appear in any of the five cases above.

%%%%%%%%%%%%%%%%%%%%%%%%%%%%%%%%%%%%%%%%%%%%%%%%%%%%
\subsection{Storage and Complexity Analysis}

\noindent{\bf Storage.} A FiRM filter entry is a bare counter, requiring
$\lceil \log_2 T_F \rceil = 11$ bits. A TinyMG entry adds a 14-bit tag
(row address within a sub-bank) and valid and lock bits to an identically
sized counter, so only 40\% of its 27 bits holds the information the
mode-switch actually consumes. Table~\ref{tab:storage} compares the two at
equal entry count: FiRM needs 352 bytes per bank\footnote{FiRM-P also needs to count the number of windows during the heavy-mode.  We avoid dedicated storage for this by repurposing the filter-counter in heavy-mode to count windows.  If the counter is at most 1250, it represents lite-mode, and activations increment the counter.  If the counter is 1251, it represents the {\em entry-window}, and if it is 1252, it represents the {\em bridge-window} (PARA with $p_1$ is used for both), and the counter is incremented only {\em once} per tREFW. If the counter is between 1253 and 2046, it represents steady-state heavy-mode, so PARA with $p_2$ is used.  If the counter reaches maximum (2047), it represents the {\em exit-window} (PARA with $p_3$ is used).  When the counter rolls over to zero, it represents lite-mode.  Thus, the default Epoch (E) for our FiRM-P implementation is 2046-1253+1 = 794 windows.} against 864 for Sigries,
a reduction by $2.5\times$, entirely from discarding metadata rather than
tracking fewer regions.

\begin{table}[htb]
\centering
\setlength{\tabcolsep}{6pt}
%\vspace{-0.1 in}
\caption{Storage cost and lookup complexity per bank ($T_{RHD}$ of 3K).}
\vspace{-0.05 in}

\label{tab:storage}
\begin{tabular}{ccc}
\toprule
 & Sigries & FiRM-P \\
\midrule
Bits per Entry & 27 (14 tag + 2 + 11) & {\bf 11} \\
Entries per Bank & 256 & 256 \\
Storage per Bank & 864 B & {\bf 352 B (0.4x)} \\
Lookup & 32-way CAM & {\bf Direct-Mapped} \\
\bottomrule
\end{tabular}
\end{table}

\noindent{\bf Complexity.} The gap in structure matters more than the gap
in bits. Every activation to a Sigries sub-bank drives a 32-way associative
match, and on a miss a second 32-way reduction to locate the minimum-count
entry for eviction. FiRM eliminates the
structure rather than shrinking it: an activation indexes the array,
performs one SRAM read-modify-write, and compares against $T_F$. No tags,
no associative match, no eviction policy.

\begin{figure*}[!htb]
    \centering
    \includegraphics[width=0.95\textwidth]{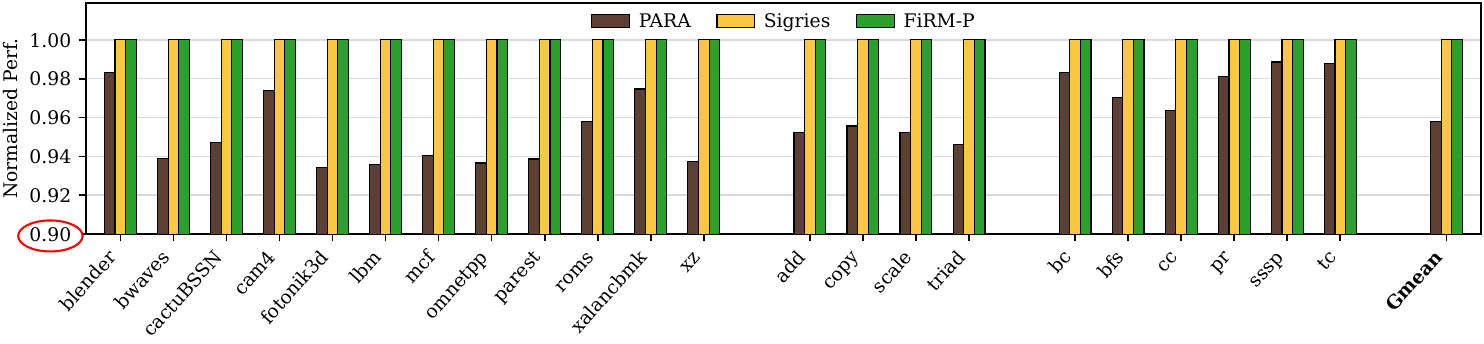}
    \vspace{-0.1 in}
    \caption{Performance of PARA, Sigries, and FiRM-P.  PARA incurs an average slowdown of 4.4\%, whereas Sigries and FiRM-P have no slowdown.}
\vspace{-0.1in}
    \label{fig:firmP_perf}
\end{figure*}

\subsection{Performance Impact for Benign Workloads}

\subsubsection{Simulation Framework}  We use DRAMSim3~\cite{li:dramsim3}, a detailed memory system simulator, to model the DDR5 configuration. Table~\ref{table:system_config} shows the configuration for our system. 

%We use the Minimalist Open Page (MOP)~\cite{kaseridis2011minimalist} policy as it performs the best for our configuration. We use an adaptive page-closure policy that closes the row if there are no pending read requests to the opened row. 

\begin {table}[h!]
\begin{center} 
\vspace{-0.15in}
\begin{footnotesize}
\caption{Baseline System Configuration} %\hrit{tRCD, tRC, tPRE (and specify the speed)}}
\vspace{-0.2 in}
\small
\begin{tabular}{|c|c|}
\hline
  Out-of-Order Cores             & 8 cores at 4GHz, 4-wide, ROB-256      \\
%  ROB size                       & 256       \\
  Last Level Cache (Shared)      & 16MB, 16-Way, 64B lines, LRU \\ \hline
 % Memory size                    & 32GB -- DDR5 \\
  Memory bus speed               & 3 GHz (6000 MT/s) \\
  Channels x Sub-Channels        & 1 (one 32GB DIMM)$\times$2 \\  
  Banks x Ranks  x Rows   & 32$\times$1$\times$128K \\
%  tRCD -- tPRE -- tRC            & 14ns -- 14ns -- 46ns \\ 
 % tDRFMsb, tDRFMab               & 240 ns and 280 ns \\ \hline
  Page Closure  & Adaptive Close Page \\
  Address Mapping & Minimalist Open Page (MOP)~\cite{kaseridis2011minimalist} \\\hline
\end{tabular}
\label{table:system_config}
\vspace{-0.05 in}
\end{footnotesize}
\end{center}
\end{table}

\subsubsection{Workload Characterization}
\label{sec:wc_characterization}
We use twelve high-MPKI benchmarks from SPEC2017~\cite{SPEC2017}, six from GAP~\cite{GAP} and four from STREAM~\cite{mccalpin1995memory}. We run applications in 8-core rate mode for 250 million instructions each. We use weighted speedup as our metric. Table~\ref{table:wc} shows our workload characteristics. The average activations per bank is 14.8 per-tREFI, so 121K per tREFW. Given 256 entries, each filter entry of FiRM-P receives an average of 474 activations per tREFW, and the most active entry we observe stays well below our $T_F$ of 1250.

\ignore{
, including the maximum activations received by any row in tREFW.  For our workloads, even the most stressful row receives only a couple of hundred activations per tREFW, consistent with Sigries. The average activation count per bank  is 13.6 per-tREFI, which is 111K per tREFW. Thus, each of the 256 filter entry of FiRM-P receives an average of 435 activations, well below our $T_F$ of 1250.
}

%, hence benign applications remain in lite-mode. 

%For example, the worst row is for parest and encounters 322 activations (our results are similar to Sigries characterization, which had the hottest with 583 activations and the second with 326 activations). 

%The average activation count per bank  is 13.6 per-tREFI, which translates to 111K per tREFW. If these activations were spread across 256 filter-entries, then each filter entry would get 435 activations on average, which is well below our $T_F$ of 1250, hence benign applications remain in lite-mode. 

\begin{table}[!htb]
  \centering
  \footnotesize
  \vspace{-0.05 in}
  \caption{Workload Characteristics.}
    \vspace{-0.1 in}
  \label{table:wc}
  \begin{tabular}{|c|c|c|c|}
    \hline
Workload    & MPKI & Avg. ACTs/Bank &	 Max ACT/Row\\
            &  & (per tREFI) &  (per tREFW)\\ \hline \hline
% bc & 59 & 12.7 & 218 \\
% bfs & 30.9 & 11.2 & 19 \\
% cc & 58.5 & 18.5 & 128 \\
% pr & 57.7 & 12 & 17 \\
% sssp & 27.4 & 11.8 & 35 \\
% tc & 87.8 & 12.2 & 18 \\ \hline
% blender & 1.5 & 5.8 & 130 \\
% bwaves & 41.6 & 15.5 & 40 \\
% cactuBSSN & 3.5 & 14 & 91 \\
% cam4 & 3.8 & 8.3 & 28 \\
% fotonik3d & 26.7 & 18.6 & 105 \\
% lbm & 27.7 & 19.7 & 92 \\
% mcf & 22.4 & 18.5 & 166 \\
% omnetpp & 10.1 & 16.8 & 286 \\
% parest & 28.9 & 14.9 & 322 \\
% roms & 9.8 & 11 & 276 \\
% xalancbmk & 1.6 & 7.5 & 84 \\
% xz & 6 & 17.1 & 217 \\ \hline
% add & 62.5 & 13.9 & 14 \\
% copy & 50 & 13.3 & 12 \\
% scale & 41.7 & 13 & 12 \\
% triad & 53.6 & 13.6 & 13 \\ \hline \hline
% Average & 32.4 & 13.6 & 105.6 \\ \hline
bc & 58.8 & 13 & 127 \\
bfs & 30.9 & 12.8 & 19 \\
cc & 57.9 & 18.4 & 63 \\
pr & 57.7 & 12.9 & 19 \\
sssp & 27.2 & 11.9 & 28 \\
tc & 87.8 & 12.4 & 20 \\ \hline
blender & 1.1 & 5 & 28 \\
bwaves & 41.6 & 17.1 & 39 \\
cactuBSSN & 3.5 & 14.6 & 50 \\
cam4 & 3.7 & 9.7 & 36 \\
fotonik3d & 26.6 & 20.3 & 75 \\
lbm & 27.7 & 20.3 & 75 \\
mcf & 18.9 & 19.9 & 101 \\
omnetpp & 9.2 & 17.4 & 139 \\
parest & 26.6 & 17.4 & 312 \\
roms & 7.9 & 12.6 & 129 \\
xalancbmk & 1.6 & 8.2 & 39 \\
xz & 5.2 & 17.7 & 153 \\ \hline
add & 62.5 & 16.7 & 15 \\
copy & 50 & 16.1 & 18 \\
scale & 41.7 & 15.9 & 15 \\
triad & 53.6 & 16.8 & 17 \\ \hline \hline
Average & 31.9 & 14.9 & 69 \\ \hline
  \end{tabular}
\vspace{-0.1 in}
\end{table}

\subsubsection{Results} Figure~\ref{fig:firmP_perf} shows the performance of PARA, Sigries, and FiRM-P, normalized to the baseline with no mitigation.  All designs use DRFMsb (with DREAM).  PARA incurs an average slowdown of 4.4\%, violating requirement R1. Both Sigries and FiRM-P always remain in lite-mode, so they incur 0\% slowdown for all workloads.

\subsection{Slowdown Under Attacks}

\begin{figure}[!htb]
    \centering
    \vspace{-0.1 in}
    \includegraphics[width=\linewidth]{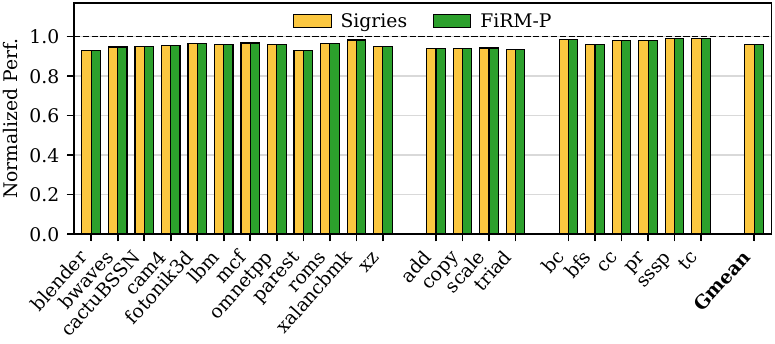}
    \vspace{-0.2 in}
    \caption{Performance impact of mitigation of Sigries and FiRM-P under attacks. Both Sigries and FiRM-P have similar slowdowns under attack.}
\vspace{-0.1 in}
    \label{fig:attackperf}
\end{figure}

We now analyze the behavior of Sigries and FiRM-P under a Rowhammer attack that targets all banks. The attacker can easily bypass the filtering of both Sigries and FiRM-P, so this reduces to their performance in heavy-mode. Figure~\ref{fig:attackperf} shows the performance of Sigries and FiRM-P under attacks. Both schemes have similar behavior as they both degenerate into PARA (with $p=1/150$). The average slowdown is 4.3\% and the worst-case slowdown is 7.7\%. We note that our slowdown for Sigries under attack is lower than reported in their paper, because Sigries paper used Naive-DRFM and we use DREAM.

\subsection{Configurability to Rowhammer Threshold}

While we design FiRM-P for $T_{RHD}$ of 3K, we can use the same hardware and configure the parameters to target different $T_{RHD}$. Table~\ref{tab:flex} shows the parameters for FiRM-P and slowdown for $T_{RHD}$ of 2K/3K/4K. When FiRM-P is operated at lower than designed $T_{RHD}$, it can incur slowdown (e.g. 0.5\% at $T_{RHD}$=2K).  Thus, FiRM-P is configurable.

\begin{table}[!htb]
\centering
\vspace{-0.05 in}

\caption{FiRM-P parameters and slowdown for different $T_{RHD}$.}
\vspace{-0.05 in}
\begin{tabular}{c|cccc|c}
\hline
\textbf{$T_{RHD}$} & \textbf{$T_F$} & \textbf{$p_1$} & \textbf{$p_2$} & \textbf{$p_3$} & \textbf{Avg. Slowdown} \\ \hline
\hline
2000 & 750  & 1/25 & 1/100 & 1/37 & 0.5\% \\
3000 & 1250 & 1/25 & 1/150 & 1/60 & 0\%   \\
4000 & 1500 & 1/50 & 1/200 & 1/75 & 0\%   \\
\hline
\end{tabular}
\vspace{-0.1 in}
\label{tab:flex}
\end{table}

\section{FiRM-D: Deterministic Filtered Mitigation}

Thus far, all the fallback options we have considered (for both Sigries and FiRM-P) have been probabilistic.  Such fallback bounds the overall design to have (a) probabilistic security and (b) reliance on a {\em Random Number Generator (RNG)}, which can be a source of both complexity and vulnerability~\cite{soothsayer}.  In this section, we show that a filtered Rowhammer mitigation can be built to be fully deterministic and at less than two-thirds the storage overhead of Sigries (and without the CAM complexity).  The key idea behind our solution, {\em FiRM-D (Deterministic)}, is to maintain a counter over a large number of rows and modulate the mitigations such that all rows are refreshed within $T_{RHD}$ activations to the region.  The mitigation rate need not be uniform, and can be modulated to be zero below a filtering threshold and higher thereafter.

\subsection{The FiRM-D Substrate: Shared Counters Across Banks}

\begin{figure}[!htb]
    \centering
    \vspace{-0.05 in}
\includegraphics[width=2.75 in]{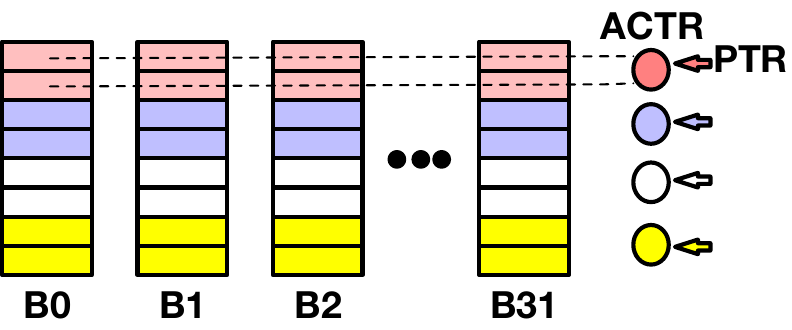}
    \caption{Substrate for FiRM-D:  All the banks share the same activation counter (ACTR).  To reduce storage, multiple rows from a bank share the ACTR. For mitigation, DRFM$_{ab}$ is used that mitigates one row from all 32 banks. For $V$ rows per bank, we need $V$ DRFM$_{ab}$ for mitigation. A Rotation Pointer (PTR) per gang identifies which row within the gang to mitigate next. }
    \label{fig:firmd}
    \vspace{-0.05 in}
\end{figure}

The enabler for the deterministic mitigation is the all-bank variant of DRFM, called DRFM$_{ab}$, which mitigates
one sampled row in {\em every} bank of the sub-channel. A design that spreads the
rows sharing a counter {\em across} banks rather than within one bank can thus
mitigate 32 of them per DRFM$_{ab}$~\cite{dream}. Figure~\ref{fig:firmd} shows the overview of such a design. Each activation counter (ACTR) is shared by a {\em gang} containing $V$ rows from every bank (Figure~\ref{fig:firmd} shows V=2). Given our budget of 256 counters per bank, we have $256 \times 32 = 8K$ counters across all 32 banks. This means, for our design, $V$ must be set to 16, given each bank has 128K rows. An activation to any row in the gang increments the ACTR of the gang. The mitigation of the gang can be accomplished by sending $V$ DRFM$_{ab}$ commands (with appropriate rows sampled for each bank). This can be done either in bulk or gradually (once per few activations).  To support the gradual schemes, each gang also has a {\em Rotation Pointer (PTR)}, which identifies the row within the gang to be refreshed next. At every tREFW, the ACTR is reset (the mitigation rate must account for loss due to reset), however, the PTR is preserved (to avoid starvation of the last row within the gang).\footnote{While Figure~\ref{fig:firmd} shows that the same bank offset from all the banks maps to one ACTR, this is inefficient if a hot page is spread across 32 banks at the same offset, leading to frequent overflow of ACTR. To avoid this, we use rows with different offset in each bank to occupy the same gang, using a per-bank xor mask, similar to the scheme used in DREAM-C~\cite{dream}.}

\subsection{FiRM-D Modulation Knob: The Rate of Mitigation}

Our main control knob is how the $V$ mitigations are performed for the gang.  We consider the three options:

\vspace{0.05 in}

\noindent{\bf DREAM-C (Bursty Mitigations):} DREAM-C is
trigger-based. When the counter reaches $T_{RHD}/2$, the whole gang is mitigated at once with  $V$
back-to-back DRFM$_{ab}$. Unfortunately, this can stall the channel for several microseconds (6.6 $\mu$s with V=16).  The microsecond-scale delay violates the liveness requirement (R3). The advantage of DREAM-C is that it has zero mitigations if the ACTR remains below $T_{RHD}/2$. We note that, as DREAM-C performs the mitigation of the entire gang in bulk, it does not need the per-gang PTR. 

\vspace{0.05 in}
\noindent{\bf Gradual (Paced Mitigation):} The burstiness of mitigation can be
amortized by pacing: issue one round every $X$ activations to a gang and advance
the rotation pointer (PTR), so successive rounds cover successive rows within the gang. Pacing also cuts the
reset tax to at most $X$ instead of $T_{RHD}/2$.  Thus, one DRFM$_{ab}$ must be sent every $X = T_{RHD}/(V+1) = 176$ activations.  At every tREFW, while we reset the ACTR, we must preserve the rotation pointer (PTR),  otherwise, the last row in the gang will get starved, leading to security violations.  The problem with the Gradual mitigation is that the paced mitigation rate applies at {\em all} times, even if the application is not under an attack, incurring an average slowdown of 38\%.

\vspace{0.05 in}
\noindent{\bf{FiRM-D (Threshold-Based Mitigation):}} Similar to FiRM-P, FiRM-D splits the activation budget into two parts: the first is a filtering threshold $T_F$, below which no mitigation is applied, and the remaining budget ($T_{RHD}-2\cdot T_F$) is split across the $V$ mitigations. The factor of $2\cdot T_F$ exists because the filter is reset every tREFW, potentially forgetting up to $T_F$ activations.  Similar to the gradual scheme, at every tREFW, while we reset the ACTR, we must preserve the rotation pointer (PTR), otherwise it can lead to security violations for the last row. 

%The two advantages of FiRM-D are: (1) it avoids doing any mitigations for benign applications (much lower activation counts per gang than $T_F$) and (2) it avoids the burstiness of DREAM-C, and limits the stall time to one DRFM$_{ab}$. 

\vspace{0.05 in}

\begin{figure}[!htb]
    \centering
    \vspace{-0.3 in}
\includegraphics[width=3.8 in]{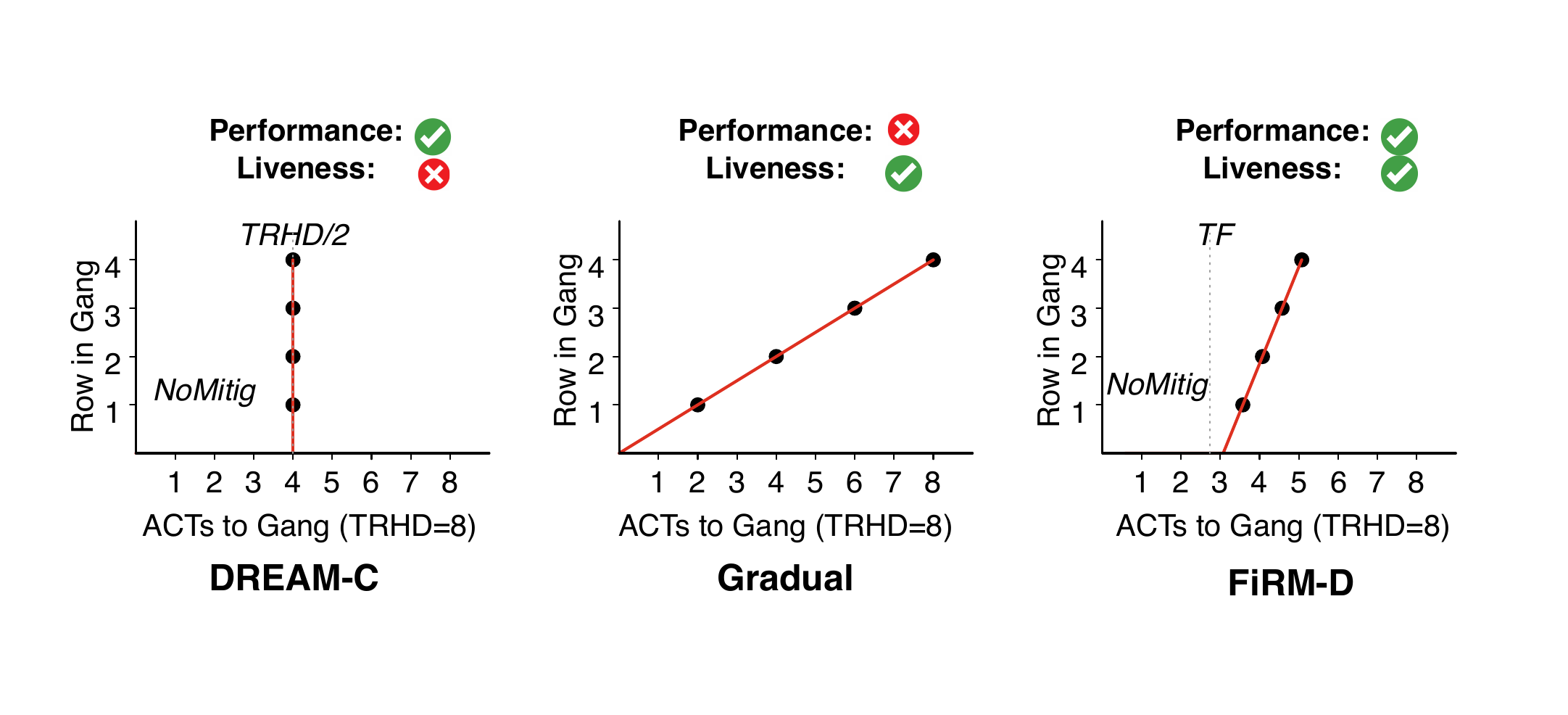}
\vspace{-0.4 in}
    \caption{Three schemes for a gang with V=4 and $T_{RHD}$ of 8. DREAM-C is bursty, Gradual has mitigations even for benign workloads, FiRM-D has filtering (no mitigations until $T_F$ and paced mitigations thereafter).  }
    \label{fig:rate}
    \vspace{-0.05 in}
\end{figure}

Figure~\ref{fig:rate} shows an overview of the three schemes, for a gang containing four rows from each bank. DREAM-C avoids mitigations until $T_{RHD}/2$ but then causes a burst of four mitigations in bulk.  The Gradual scheme pays constant overheads even when the system is not under an attack, and this may be unacceptably high for benign workloads. FiRM-D balances both: it avoids mitigations until $T_F$, then performs paced mitigation, albeit at a faster rate than the Gradual design.

\begin{figure*}[!htb]
    \centering
    \includegraphics[width=\linewidth]{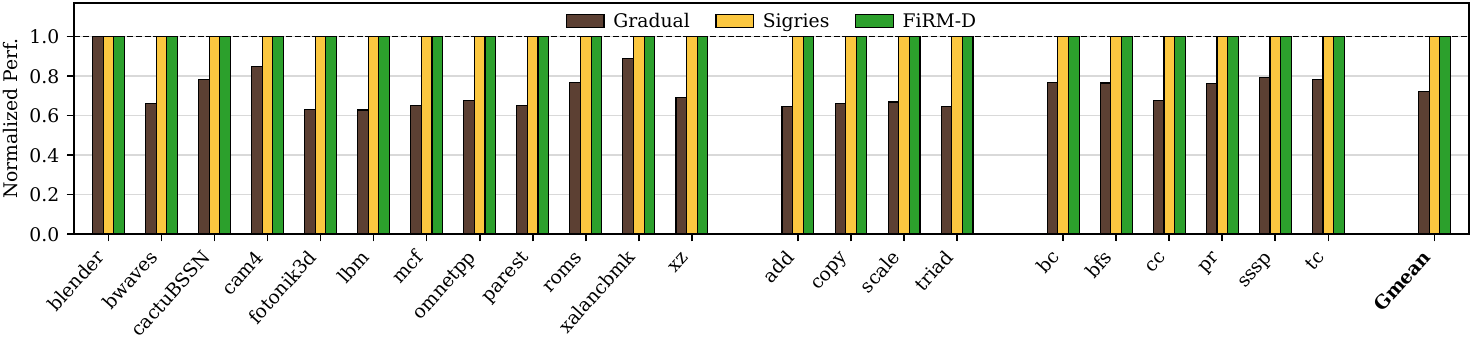}
    \vspace{-0.2 in}
    \caption{Performance of Gradual, Sigries, and FiRM-D. Gradual incurs an average slowdown of 38\%, whereas Sigries and FiRM-D have zero slowdown.}
%\vspace{-0.15 in}
    \label{fig:firmd_perf}
\end{figure*}

\subsection{Parameters}

All three schemes can be described with two parameters: $T_F$ and $APD$ (ACTs per DRFM).  Table~\ref{tab:firmdparams} shows the parameters for the three schemes for our target $T_{RHD}$ of 3K.  We assume a gang size containing V=16 rows from each bank. 

\begin{table}[htb]
\centering
\vspace{-0.1in}
\caption{Deterministic mitigation options at $T_{RHD}$ of 3K.}
\vspace{-0.05in}
\label{tab:firmdparams}
\footnotesize
\setlength{\tabcolsep}{4pt}
\begin{tabular}{lcccc}
\toprule
Scheme & Gang Size & Filter Th. & ACTs per & Stall per \\
       & ($V$)     & ($T_F$)    & DRFM (APD) & Mitig. (ns) \\
\midrule
DREAM-C & 16 & 1500 & 0 (for 16)  & 6600 \\
Gradual & 16 & 0    & 176 & 411  \\
\textbf{FiRM-D} & \textbf{16} & \textbf{1250} & \textbf{29} & \textbf{411} \\
\bottomrule
\end{tabular}
%\vspace{-0.1in}
\end{table}

\subsection{Bounding Performance Under Attack}

\begin{figure}[!htb]
    \centering
    \vspace{-0.05 in}
\includegraphics[width=3.5 in]{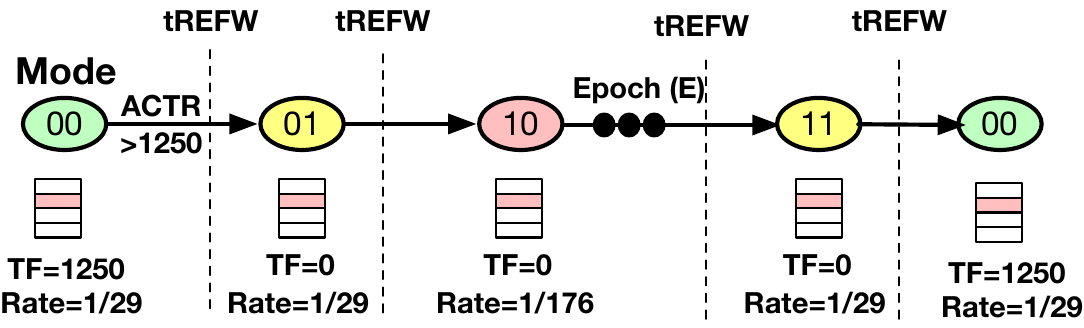}
   %  \vspace{-0.05 in}
    \caption{Epoch-Based FiRM-D (enables better performance under attacks)}
    \label{fig:firmdepoch}
    \vspace{-0.15 in}
\end{figure}

We note that, once the filtering threshold is breached, the mitigation rate of FiRM-D is 6x higher (1-per-29 vs. 1-per-176) than the Gradual scheme.  This occurs because the Gradual scheme spreads its 16 mitigations over the entire $T_{RHD}$, however, FiRM-D must squeeze its 16 mitigations within a much shorter budget of $T_{RHD}-2 \cdot T_F$. Thus, under attacks, the Gradual scheme would have 6x lower slowdowns than FiRM-D.  To achieve the best of both worlds (zero slowdown for benign applications and better bounds under attacks), we take an epoch-based approach, similar to FiRM-P. Figure~\ref{fig:firmdepoch} shows the overview of our FiRM-D-Epoch scheme.  It remains in FiRM-D (Mode-00) until the ACTR is below $T_F$.  Then, at the end of tREFW, it moves to a {\em{fast}} Gradual scheme (Mode-01) that uses a rate of 1-per-29 ACT for secure transition and lasts for exactly one tREFW. Then, it moves to a  Gradual scheme (Mode-10) with a rate of 1-per-176 and lasts for E windows (hundreds of tREFW). Then, before transition, it spends one more tREFW in {\em{fast}} Gradual scheme (Mode-11), before finally transitioning back to FiRM-D (Mode-00).

\newpage

\begin{figure*}[t]
    \centering
\includegraphics[width=\textwidth]{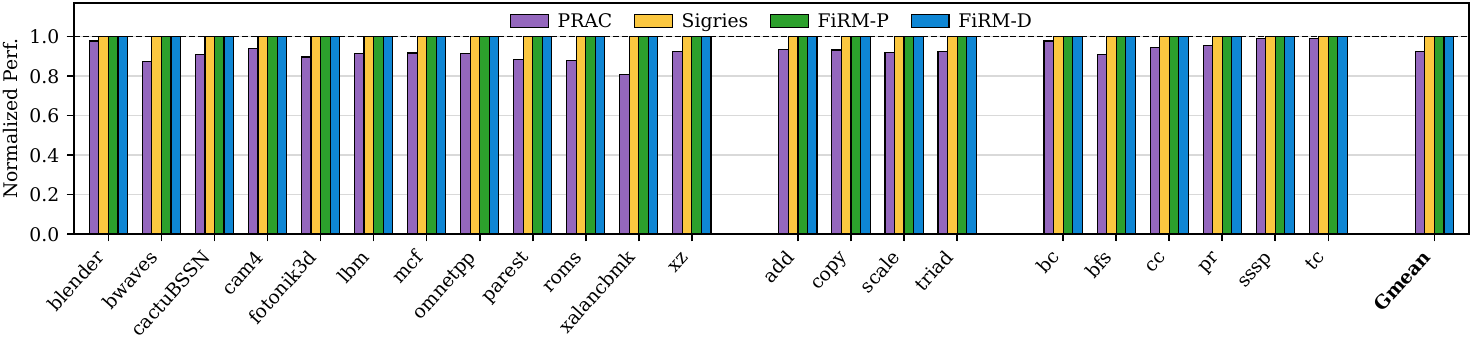}

\vspace{-0.15 in}
    \caption{Performance of PRAC, Sigries, FiRM-P and FiRM-D. PRAC has an average slowdown of 8.4\%, whereas the other three designs have zero slowdown.}
\vspace{-0.15 in}
    \label{fig:prac}
\end{figure*}

% ============================================================
% Section V-E: Security Analysis (FiRM-D) -- corrected, ~30 lines
% (previous version ~35 lines; saves ~5 to pull Section G onto page 11)
% Changes: refresh-interval framing restored in the intro; bound restored in
% Case 4; 3919; floor in Case 1; Case 5 folded into Case 2; grammar fixes.
% ============================================================

\subsection{Security Analysis}

Security requires that no row receives more than $T_{RHD}$ activations
between two refreshes of the victim. Two facts simplify the argument. First,
as the ACTR reset is not aligned with the refresh of a row, a refresh
interval spans at most one reset and at most one mode transition. Second,
the PTR is preserved across resets and transitions, so a row is served once
every $V$ mitigations: at one mitigation per $X$ activations, a row waits at
most $V \cdot X$ activations for its turn, plus the up-to $X-1$ activations
of pacing progress discarded at the reset. An interval granting a filtering
allowance of $A$ is thus secure if $A + (V{+}1) \cdot X \leq T_{RHD}$, with
$X$ the slowest rate in force.

\vspace{0.03 in}
\noindent\textbf{Case 1: Mode-00 (FiRM-D).} An interval straddling a reset
receives an allowance of $T_F$ on either side, so $A = 2 \cdot T_F = 2500$
and the rate must satisfy
$X \leq \lfloor (T_{RHD} - 2 \cdot T_F)/(V{+}1) \rfloor = 29$, which is
exactly the FiRM-D rate (worst case 2993).

\vspace{0.03 in}

\noindent\textbf{Case 2: Gradual (Mode-01, 10, 11).} Filtering is disabled,
so $A = 0$ and the entire budget is available to pacing: the rate may be as
slow as $T_{RHD}/(V{+}1) = 176$, which is the Mode-10 rate (worst case
2992). Mode-01 and Mode-11 are faster, and the two transitions Mode-01 to Mode-10 and
Mode-10 to Mode-11 also charge no allowance, so this case covers them.

\vspace{0.03 in}

\noindent\textbf{Case 3: Entry (Mode-00 to Mode-01).} The transition
coincides with a reset, so the interval charges only one allowance,
$A = T_F$, leaving room for a rate no slower than
$(T_{RHD} - T_F)/(V{+}1) = 102$. Mode-01, at 1-per-29, is secure; entering
Mode-10 directly would not be, as the row could accrue
$T_F + 17 \cdot 176 = 4242$ activations.

\vspace{0.03 in}

\noindent\textbf{Case 4: Exit (Mode-11 to Mode-00).} This is the mirror of
Case 3. At the Mode-10 rate, the last row of the gang
could instead wait $15 \cdot 176$ activations for its turn in the exit
window and a further $T_F + 29$ after the reset, for a total of 3919. Operating at 
Mode-11 at a rate of 1-per-29 removes this case.

\subsection{Performance Impact on Benign Workloads}

Figure~\ref{fig:firmd_perf} shows the performance of Gradual, FiRM-D, and Sigries normalized to the baseline without any mitigation.  For FiRM-D, no workload in our suite drives an entry past $T_F$ (the average ACTR reaches 474 activations, whereas $T_F$=1250), so we expect  FiRM-D to issue no mitigations.  The Gradual scheme suffers an average slowdown of 38\% and up to 60\%, whereas both FiRM-D and Sigries have zero slowdown.

%\newpage
\subsection{Storage Analysis}

Each entry for FiRM-D requires an ACTR (11-bit), a Rotation-Pointer
(4-bit), and Mode (2-bit). To reduce storage, we use a single global Epoch counter. Table~\ref{tab:storage-all} compares the
storage of FiRM-P, FiRM-D, and Sigries. FiRM-P requires less than half the storage of Sigries, whereas FiRM-D requires less than two-thirds the storage of  Sigries. Both FiRM-P
and FiRM-D avoid the CAM complexity of Sigries. 

%FiRM-D additionally removes the need for a random number generator, which can be a source of complexity and vulnerability.

%Each entry for FiRM-D requires an ACTR (11-bit),  a Rotation-Pointer (4-bit), and Mode (2-bit).  Table~\ref{tab:storage-all} compares the storage of FiRM-P, FiRM-D, and Sigries. FiRM-D uses
%$1.5\times$ less storage than Sigries. Furthermore, both FiRM-P and FiRM-D avoids the CAM complexity of Sigries.  FiRM-D additionally removes the need for random number generator.

\begin{table}[h]
\centering
\vspace{-0.1 in}
\caption{Storage and Complexity per Bank ($T_{RHD}$ of 3K).}
\vspace{-0.05 in}

\label{tab:storage-all}
\begin{tabular}{lccc}
\hline
 & \textbf{Sigries} & \textbf{FiRM-P} & \textbf{FiRM-D} \\
\hline \hline
Bits per Entry   & 27 (14+2+11) & 11 & 17 (11+4+2) \\
Entries per Bank & 256 & 256 & 256 \\
Storage per Bank & 864 B & 352 B (0.4$\times$) & 544 B (0.63$\times$) \\
Lookup           & 32-way CAM & Direct-Mapped & Direct-Mapped \\
Requires RNG     & Yes & Yes & No \\
Security         & Probabilistic & Probabilistic & Deterministic \\
\hline
\vspace{-0.1 in}
\end{tabular}
\end{table}

\subsection{Performance Impact Under Attacks}

%While the filtering threshold of FiRM-D is sufficient to handle benign workloads, an attacker could easily bypass the filtering threshold by continuously activating the rows that belong to the same gang. 

%Once the filter is bypassed, FiRM-D degenerates in the Gradual scheme (with a faster rate for the first tREW, but the same rate as the Gradual scheme for E epochs). 

%Figure~\ref{fig:firmd_attack} shows the performance of FiRM-D, FiRM-P, and Sigries for the benign applications in the presence of an attack that tries to cause mitigations at the maximum rate.  While both FiRM-P and Sigries have an average slowdown of 4.4\% under attack, FiRM-D has an average slowdown of 95\%.  The higher slowdown is expected as FiRM-D must refresh the entire region (instead of only one aggressor row). %We note that the overhead is graceful rather than a microsecond-level stall, so liveness is preserved (R3). 

%Figure~\ref{fig:firmd_attack} shows the performance of benign workloads when a co-running attacker causes mitigations at the maximum rate. Sigries and FiRM-P incur an average slowdown of 4.3\%, and FiRM-D incurs 2$\times$. Unlike Sigries and FiRM-P, which mitigate one row, FiRM-D refreshes the entire gang, so it incurs more mitigations. However, FiRM-D has the advantage of deterministic security.  We note the slowdown for FiRM-D under attack is lower than other memory contention attacks, such as row-buffer conflicts~\cite{DRAMA,MEMDOS}. Also, the overhead is graceful rather than a microsecond-scale stall, so liveness (R3) is preserved, and it lasts only as long as the attack.
 
\begin{figure}[!htb]
    \centering
   % \vspace{-0.1 in}
    \includegraphics[width=\linewidth]{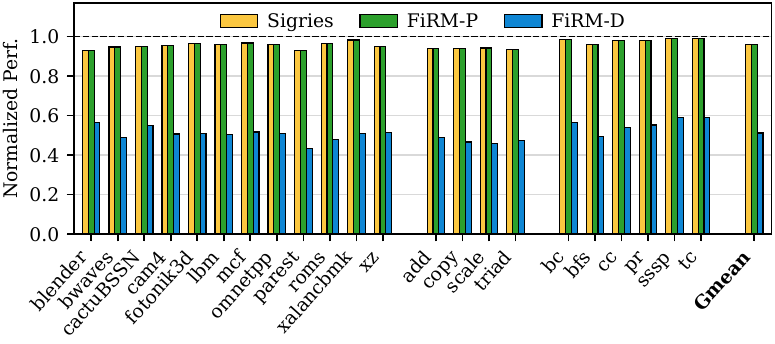}
    \vspace{-0.1 in}
    \caption{Performance impact of mitigation of Sigries, FiRM-P and FiRM-D under an attack with maximum mitigation rate.}
\vspace{-0.1 in}
    \label{fig:firmd_attack}
\end{figure}

Figure~\ref{fig:firmd_attack} shows the performance of benign workloads when
a co-running attacker causes mitigations at the maximum rate. Sigries and
FiRM-P incur an average slowdown of 4.3\%, whereas FiRM-D incurs 2$\times$.
The gap has two sources: (1) FiRM-D must refresh all the rows in the gang (using $V$ mitigations), whereas FiRM-P does one mitigation to a row. (2) Each mitigation stalls all 32
banks, so benign co-runners absorb stalls even on banks the attacker did not
mitigate.

We argue this cost is acceptable. The rate is bounded and known at design
time (at most one DRFM$_{ab}$ per 29 activations), so FiRM-D has no
performance tail and meets R2. The longest stall is a single DRFM$_{ab}$
(411~ns), $16\times$ below DREAM-C and well under the microsecond scale that
R3 prohibits. The degradation is within the range of what a co-located tenant can already
inflict through memory contention alone~\cite{MEMDOS,DRAMA}, and it ceases with the
attack. FiRM-D thus trades attack-time throughput for deterministic security.

%\newpage

\section{Related Work}

\noindent\textbf{PRAC:} Recent JEDEC specification introduced \emph{Per-Row Activation Counting (PRAC)}, which maintains an activation count
alongside every row~\cite{qureshi2024moat,woo2025qprac,JEDEC-PRAC}. To enable counter update, PRAC increases $t_{RC}$, so it incurs a slowdown even when the system is not under
attack. Figure~\ref{fig:prac} compares PRAC with Sigries, FiRM-P, and
FiRM-D at $T_{RHD}$ of 3K. PRAC suffers 8.4\% on average, whereas all three
MC-side designs are free on benign workloads. Furthermore, PRAC remains optional for DDR5, whereas the MC-side designs can protect DDR5 DIMMs (that may remain in use over the next decade).

\vspace{0.05 in}
\noindent\textbf{Filtered Mitigation In-DRAM:} Our design of FiRM-P is inspired by MIRZA~\cite{taneja2026mirza}, an  in-DRAM mitigation that uses filtering and randomization. As MIRZA is
in-DRAM, its filter reset is coordinated with refresh, so it does not suffer from security issues in transitioning from one mode to another. As FiRM is MC-based, it has no visibility into when a row is refreshed, which causes both the filter budget to be doubled and security issues in transitioning from randomization to filtering. Finally, FiRM-P is carefully designed to provide a slowdown of 4.3\% even under attacks.

\ignore{
\vspace{0.05 in}

\noindent\textbf{Counter-Based Trackers at the MC:} Graphene~\cite{park2020graphene} uses
Misra-Gries to identify the top-$K$ rows, but needs a CAM with hundreds of entries per bank.
Hydra~\cite{hydra} and START~\cite{start} lower this cost by keeping
per-row counters outside dedicated SRAM, in DRAM and in the LLC
respectively. 
}

%\vspace{0.05 in}

%\noindent\textbf{Alternative Mitigative Actions:}  Other lines of work change the mitigating action: instead of using victim-refresh, they  either migrate aggressor rows~\cite{AQUA, saileshwar2022RRS} or rate-limit accesses to them~\cite{yauglikcci2021blockhammer}. However, these schemes incur multi-microsecond delays and hence do not obey the liveness property. 

\ignore{
\vspace{0.05 in}

\noindent\textbf{Auditing Deployed Defenses:} A recurring result in this
area is that a defense which is sound in the abstract fails as deployed.
TRRespass~\cite{1} and Blacksmith~\cite{2} broke the in-DRAM TRR shipped in
DDR4. Our analysis of Sigries
continues this line for MC-side defenses, and finds that the failure comes
not from either mode but from their composition.
}

\section{Conclusion}

Sigries is an MC-side Rowhammer defense deployed in production. We show a new vulnerability in Sigries. We also show that by targeting multiple sub-banks in a Round-Robin manner, an attacker can keep a vulnerability window open at almost 
every instant, reducing the system MTTF by eight orders of magnitude. We propose {\em FiRM}, a principled filtered mitigation. FiRM partitions the activation budget across the filter
and the fallback to ensure security across modes and 
transitions.
FiRM-P uses non-uniform sampling and needs less than half the storage of
Sigries. FiRM-D provides
deterministic security and needs less than two-thirds the storage of Sigries. Both designs incur zero slowdown for
benign workloads. We show that neither the security tradeoff nor the CAM complexity of
Sigries is essential for a practical mitigation.

\newpage
 
\appendices
\section{On Optimizing for the Blind-Static Attacker}
\label{app:blind-static}
 
Both Sigries and FiRM assume an attacker who knows the mitigation
algorithm and its parameters (Section~\ref{sec:threat}), therefore, neither
design seeks to obtain security through obscurity. Accordingly, we evaluate both the security properties and performance under attack, assuming that the attacker is aware of the algorithm and is trying to cause harm (either failures or slowdown) by crafting the most stressful access pattern based on the knowledge of the underlying defense.  

One may argue that the Misra-Gries tracker is more precise than the group-counting for FiRM-P and can provide performance benefits for specifically crafted weaker attack patterns.  However, this is a regime which requires the opposite assumption -- the attacker is unaware of the defense parameters and is also unwilling to change the attack pattern.  The tracker outperforms the filter only on
patterns that hammer fewer rows than it has entries, and an attacker who
knows the design has no incentive to produce such a pattern. We contend that the
precise tracker of Sigries therefore optimizes for a regime that is not meaningful in practice: benign workloads never enter it, and an attacker who may try to launch such patterns has nothing to gain (failures or slowdown) and hence is unlikely to do it for any sustained period of time.

\vspace{0.05 in}
 
\noindent\textbf{The Regime.} Let $C$ denote the number of entries in a
TinyMG ($C{=}32$ in our configuration). A pattern that hammers at most
 $C$ distinct rows within a sub-bank is tracked exactly: the
spill-counter never reaches $T_{MG}$, Sigries never leaves lite-mode,
and a tracked row is mitigated once per $T_{MG}{=}1500$ activations.
FiRM has no notion of \emph{how many} rows are hot, as the filter
counter is a per-region activation total, so any pattern that drives a
region past $T_F$ trips the fallback. On a 10-row pattern
(\emph{decahammer}), Sigries therefore would issue $10\times$ fewer
mitigations than FiRM-P (1-in-1500 vs. 1-in-150).

\vspace{0.05 in}

\noindent\textbf{Not for Benign Workloads.} Table~\ref{table:wc} shows
that the most active row across our 22 workloads receives 312
activations per $t_{REFW}$, and the busiest FiRM filter entry stays well
below $T_F{=}1250$. Neither design issues a single mitigation on this
suite. Thus, the precise counting of TinyMG is not meant to mitigate the hot rows in benign applications (e.g., the hottest row in the Sigries study~\cite{sigries} had 583 activations, well below the $T_{MG}$ of 1500). So, the differentiator between the tracker and the filter is visible only under deliberate hammering.

\vspace{0.05 in}

\noindent\textbf{Not Attackers With Anything to Gain.} An attacker has an incentive to 
persist only if the attack yields something: a bit-flip, or meaningful
slowdown of the system. Decahammer against Sigries delivers
neither. It cannot flip a bit, as the tracker forces a mitigation every
$T_{MG}$ activations and the row never approaches $T_{RHD}$. It is also
strictly dominated as a denial-of-service vehicle: hammering $C{+}1$
rows instead of $C$ pushes the sub-bank into heavy-mode and yields
4.3\% slowdown rather than 0.4\%. Sigries thus wins only where the attacker deliberately settles for 
$10\times$ lower harm.

\newpage 
\noindent\textbf{A Taxonomy of Attackers.} To better understand the landscape of attacks, we classify the attackers along two axes.  First, based on the 
\emph{Knowledge}, the attacker is either \emph{Blind} (the attacker knows nothing of
the deployed mitigation) or \emph{Informed} (the attacker knows the
algorithm and its parameters). Second, based on \emph{Skill}, the attacker is either \emph{Static} (the
attacker runs only a small set of fixed patterns) or \emph{Adaptive} (the
attacker varies the pattern and tries to converge on the most effective one).
Figure~\ref{fig:attacker-taxonomy} shows the four types of attackers.

\begin{figure}[h]
\centering
\begin{tikzpicture}[font=\footnotesize]
  \def\cw{2.75}   % cell width
  \def\ch{1.15}   % cell height
 
  % --- Blind-Static: the only cell where the Sigries tracker helps ---
  \fill[advpink, draw=advpinkline, line width=0.6pt]
        (0,0) rectangle ++(\cw,-\ch);
  \node[align=center] at (0.5*\cw,-0.5*\ch)
       {\bfseries Advantage\\[1pt]\scriptsize\itshape (not useful regime)};
 
  % --- Blind-Adaptive ---
  \fill[noadvgreen, draw=noadvgreenline, line width=0.6pt]
        (\cw,0) rectangle ++(\cw,-\ch);
  \node[align=center] at (1.5*\cw,-0.5*\ch)
       {No advantage\\
       [1pt]\scriptsize\itshape (valid regime)};
 
  % --- Informed-Static ---
  \fill[noadvgreen, draw=noadvgreenline, line width=0.6pt]
        (0,-\ch) rectangle ++(\cw,-\ch);
  \node[align=center] at (0.5*\cw,-1.5*\ch)
       {No advantage\\[1pt]\scriptsize\itshape (valid regime)};
 
  % --- Informed-Adaptive (our threat model) ---
  \fill[noadvgreen, draw=noadvgreenline, line width=0.6pt]
        (\cw,-\ch) rectangle ++(\cw,-\ch);
  \node[align=center] at (1.5*\cw,-1.5*\ch)
       {No advantage\\[1pt]\scriptsize\itshape (valid regime)};
 
  % --- headers ---
  \node at (0.5*\cw, 0.30) {\bfseries Static};
  \node at (1.5*\cw, 0.30) {\bfseries Adaptive};
  \node at (\cw, 0.76) {\itshape Skill};
  \node[anchor=east] at (-0.12,-0.5*\ch) {\bfseries Blind};
  \node[anchor=east] at (-0.12,-1.5*\ch) {\bfseries Informed};
  \node[rotate=90] at (-2.00,-\ch) {\itshape Knowledge};
\end{tikzpicture}
\caption{The four attacker classes. The
Misra-Gries tracker of Sigries outperforms the FiRM filter only against
a Blind-Static attacker, a regime where the attacker has no incentive to run for meaningful time.}
\label{fig:attacker-taxonomy}
\end{figure}
 
An example of \emph{Blind-Adaptive} is the fuzzing approach of TRRespass~\cite{frigo2020trrespass}
and Blacksmith~\cite{jattke2021blacksmith}, which defeat in-DRAM TRR without any
knowledge of its design by searching the space of patterns until one
works. \emph{Informed-Static} is the worst-case analysis by which
defenses such as ProTRR~\cite{protrr} and
MINT~\cite{qureshi2024mint} are evaluated: the pattern is derived
analytically from a known design and applied as given. \emph{Informed-Adaptive} begins from such a pattern and adjusts it to handle
runtime contention, for example to exploit {\em Refresh Postponement}~\cite{posthammer}. A \emph{Blind-Static} attacker in theory runs a 
few canonical patterns, for a short time, whether or not they
accomplish anything.
 
The three classes that describe reported Rowhammer attacks are thus
Blind-Adaptive, Informed-Static, and Informed-Adaptive, and in all three cells, 
Sigries and FiRM behave identically. The tracker earns its advantage
only in the one quadrant that provides no incentive to run for a long time.

\vspace{0.05 in}
 
\noindent\textbf{The Residual Case.} What remains is an
attacker who is uninformed, unadaptive, and persistent enough to run a
failing pattern long enough to matter for system performance, while gaining nothing itself. Even in this case, FiRM-P bounds the attacker to only a small slowdown in the heavy-mode (4.3\% on average and 7.7\% at worst). A Blind-Static attacker moves
FiRM from one point inside the minor slowdown envelope to another. It creates no new
worst case, R1 and R2 are unaffected, and the same 4.3\% is available
against Sigries to any attacker willing to change one constant in the
pattern.

\vspace{0.05 in}
\noindent\textbf{Implication.} The tracker buys lower slowdown only
against attacks that are already failing, and only from an attacker who
is simultaneously uninformed and unwilling to experiment. The price is a
32-way associative lookup on every activation, $2.5\times$ the storage,
and security vulnerability.  Precision is worth paying for when it improves the cases a system encounters for significant time, the Blind-Static quadrant is not one of them.

\section{On Setting the Sampling Rate for PARA}
\label{app:para-sample}

The sampling rate $p$ for PARA must be set carefully, as it determines
both the security level (the Mean-Time-to-Failure under continuous
attack) and the performance overhead (the rate of mitigations). Similar
to prior works~\cite{jaleel2024pride,qureshi2024mint,AutoRFM}, we target
an MTTF of 1 failure per 10K years per bank under continuous attack, as
this yields a bank-failure rate comparable to that of naturally occurring
errors. For this target, we set $p$ for a given $T_{RHD}$ using the
approximation of Equation~\ref{eq:para}.

\begin{equation}
\label{eq:para}
    p = 20/T_{RHD}
\end{equation}

A more precise rate can be obtained with the recurrence-based method of
Saroiu and Wolman~\cite{Sampling}. Table~\ref{tab:swpara} compares the
two, both as absolute probabilities and as coefficients relative to
$T_{RHD}$. Our approximation tracks the precise method closely, with a
coefficient of 20 against 19.02--19.60, and errs on the safe side: it
samples 2--5\% more often than required, giving PARA slightly more
chances to catch an aggressor. The same holds for the fallback of
Sigries, so the MTTF we report under the Round-Robin Attack is
conservative in this respect as well. The two methods are virtually
indistinguishable in performance, as the difference in mitigation rate
is negligible.

\begin{table}[hbt]
\centering
\vspace{0.1 in}

\caption{Sampling probability from the precise Saroiu-Wolman
computation~\cite{Sampling} and from the $20/T_{RHD}$ approximation.
Our approximation is 2--5\% more conservative.}
\label{tab:swpara}
\footnotesize
\begin{tabular}{ccccc}
\toprule
 & \multicolumn{2}{c}{Precise} & \multicolumn{2}{c}{Approximate} \\
\cmidrule(lr){2-3}\cmidrule(lr){4-5}
$T_{RHD}$ & $p$ & Coefficient & $p$ & Coefficient \\
\midrule
500  & 1/25.5  & $19.60/T_{RHD}$ & 1/25  & $20/T_{RHD}$ \\
1000 & 1/51.4  & $19.45/T_{RHD}$ & 1/50  & $20/T_{RHD}$ \\
1500 & 1/77.7  & $19.31/T_{RHD}$ & 1/75  & $20/T_{RHD}$ \\
2000 & 1/104.3 & $19.18/T_{RHD}$ & 1/100 & $20/T_{RHD}$ \\
2500 & 1/130.9 & $19.09/T_{RHD}$ & 1/125 & $20/T_{RHD}$ \\
3000 & 1/157.7 & $19.02/T_{RHD}$ & 1/150 & $20/T_{RHD}$ \\
\bottomrule
\vspace{0.1 in}
\end{tabular}
\end{table}

\section*{Responsible Disclosure}
On July 8, 2026, we contacted the Sigries authors to note that the
heavy-mode to lite-mode transition may be unsafe, as the tracker is
guaranteed to be secure only if a row enters with zero unmitigated
activations, whereas a row leaving sampling-mode may carry some. We
shared a draft of this paper with them on August 3, 2026, and have since
incorporated their feedback on the description of Sigries. Our analysis
of the Round-Robin Attack has been revised since that draft, and the
MTTF we report here is lower than the one shared with them. This work
analyzes a design as described in a published paper, using estimated
parameters. We did not attempt to mount the attack on any deployed
system.

\section*{Impact Statement}
Our analysis shows that an attacker running a Round-Robin pattern across
sub-banks can bypass the protection Sigries provides, so a system
deploying it must rely on in-DRAM mitigations to tolerate Rowhammer.
We note that this leaves such a system no worse off than the many
servers shipping today with no SoC-level mitigation at all, and that
the vulnerability arises from how the two modes are parameterized
rather than from any flaw in either mode individually. Our paper 
shows that our filtered design can retain the benefits of Sigries while
remaining secure across modes and transitions.

\section*{Acknowledgments}
We thank the Sigries team for describing their design in detail in a
research paper. Progress in hardware security depends on avoiding
security-by-obscurity, and we commend the Sigries team for opening
their design to external analysis. We thank Stefan Saroiu and Sujay
Yadalam for discussions at ISCA and for answering our questions.

% \section*{Acknowledgements}
% This document is an updated version of HPCA 2022 and 2023, which, in
% turn, has been derived from two previous conferences, in particular
% HPCA 2021 and MICRO 2021, which, in turn, are derived from past MICRO,
% HPCA, ISCA, and ASPLOS conferences.

%%%%%%% -- PAPER CONTENT ENDS -- %%%%%%%%

%\clearpage
%%%%%%%%% -- BIB STYLE AND FILE -- %%%%%%%%
\balance
\bibliographystyle{IEEEtranS}
\bibliography{refs}
%%%%%%%%%%%%%%%%%%%%%%%%%%%%%%%%%%%%

\end{document}